\documentclass[10pt,journal,compsoc]{IEEEtran}
\ifCLASSOPTIONcompsoc
  \usepackage[nocompress]{cite}
\else
  \usepackage{cite}
\fi
\ifCLASSINFOpdf
\else
\fi
\usepackage{paralist}
\usepackage{xcolor}
\usepackage[normalem]{ulem}
\usepackage{mathtools}
\usepackage{amssymb}
\usepackage{hyperref}
\usepackage{multirow}

\usepackage{graphicx}
\usepackage{array}
\usepackage{balance}
\usepackage{float}
\usepackage{enumitem}
\usepackage[normalem]{ulem}
\useunder{\uline}{\ul}{}

\newcommand{\zhihua}[2]{#2}
\newcommand{\mzhihua}[2]{#2}

\definecolor{orange}{RGB}{245, 133, 24}

\newcommand{\eg}{\textit{e.g.}}
\newcommand{\ie}{\textit{i.e.}}

\newcommand{\name}{{\textit{GNNLens}}}

\begin{document}
%

\title{{\name}: A Visual Analytics Approach for Prediction Error Diagnosis of Graph Neural Networks}

%
%
%
%


\author{Zhihua~Jin, Yong~Wang, Qianwen~Wang, Yao~Ming, Tengfei~Ma, and~Huamin~Qu,~\IEEEmembership{Member,~IEEE}

\IEEEcompsocitemizethanks{\IEEEcompsocthanksitem Z. Jin and H. Qu are with Hong Kong University of Science and
Technology, Hong Kong,
China.\protect\\
E-mail: {zjinak, huamin}@cse.ust.hk.

\IEEEcompsocthanksitem Y. Wang is with Singapore Management University. \protect\\ E-mail: yongwang@smu.edu.sg.
\IEEEcompsocthanksitem Q. Wang is with Harvard University. \protect\\ E-mail: qianwen\_wang@hms.harvard.edu.
\IEEEcompsocthanksitem Y. Ming is with Bloomberg LP. \protect\\ E-mail: yming7@bloomberg.net.
\IEEEcompsocthanksitem T. Ma is with IBM T. J. Watson Research Center. \protect\\ E-mail:
tengfei.ma1@ibm.com.
}
\thanks{Manuscript received XX XX, 2020; revised XX XX, 2020.}}

%
%

\markboth{IEEE TRANSACTIONS ON VISUALIZATION AND COMPUTER GRAPHICS,~VOL.~XX, NO.~XX, XX~2020}%
{Shell \MakeLowercase{\textit{et al.}}: Bare Demo of IEEEtran.cls for Computer Society Journals}
%




\IEEEtitleabstractindextext{%

\begin{abstract}
Graph Neural Networks (GNNs) aim to extend deep learning techniques to graph data and have achieved significant progress in graph analysis tasks (e.g., node classification) in recent years.
However, similar to other deep neural networks like Convolutional Neural Networks (CNNs) and Recurrent Neural Networks (RNNs), GNNs behave like a black box with their details hidden from model developers and users. It is therefore difficult to diagnose possible errors of GNNs. 
Despite many visual analytics studies being done on CNNs and RNNs, little research has addressed the challenges for GNNs.
This paper fills the research gap with an interactive visual analysis tool, {\name}, to assist model developers and users in understanding and analyzing GNNs.
Specifically, Parallel Sets View and Projection View enable users to quickly identify and validate error patterns in the set of wrong predictions; Graph View and Feature Matrix View offer a detailed analysis of individual nodes to assist users in forming hypotheses about the error patterns.
Since GNNs jointly model the graph structure and the node features, we reveal the relative influences of the two types of information by comparing the predictions of three models: GNN, Multi-Layer Perceptron (MLP), and GNN Without Using Features (GNNWUF).
Two case studies and interviews with domain experts demonstrate the effectiveness of {\name} in facilitating the understanding of GNN models and their errors.
\end{abstract}

\begin{IEEEkeywords}
Graph Neural Networks, Error Diagnosis, Visualization.
\end{IEEEkeywords}}

\maketitle
 
\IEEEdisplaynontitleabstractindextext

%

\begin{figure*}[htb]
\centering 
\includegraphics[width=\linewidth]{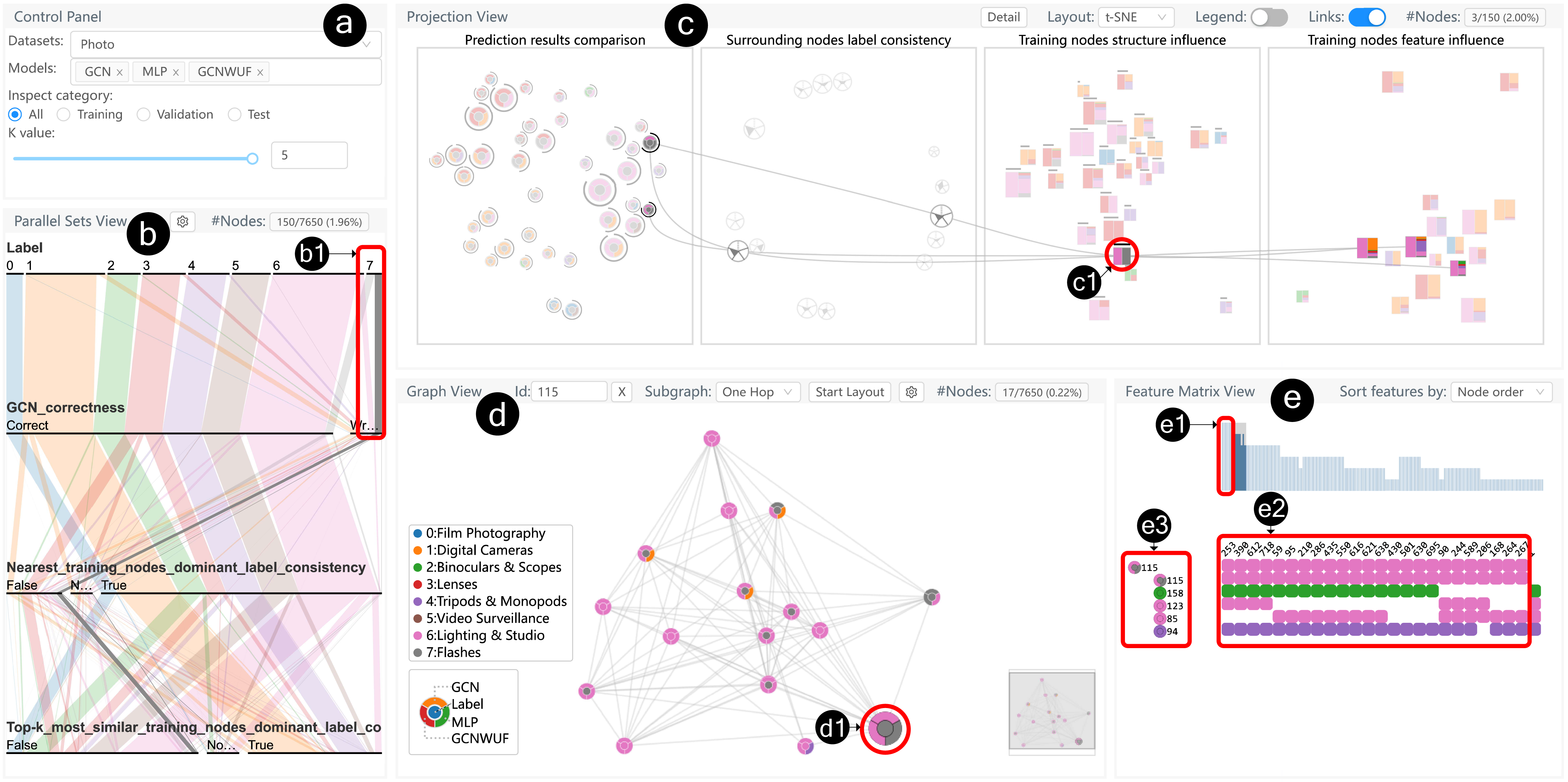}
\caption{{\name} is a visual analytics tool that helps model developers and users understand and diagnose GNNs. {\name} consists of Control Panel, Parallel Sets View, Projection View, Graph View, and Feature Matrix View.
(a) The Control Panel enables users to interactively configure the basic parameters (e.g., the dataset). (b) The Parallel Sets View displays the overall distribution of node properties. 
(c) The Projection View further visualizes four groups of key metrics of nodes that have a direct influence on GNN model performances. Interactions such as lasso-selection and linked highlighting are also supported to facilitate convenient exploration between the GNN prediction results on individual nodes and node properties.
(d) The Graph View provides users with the interactive exploration of the detailed topology structures around the node of user interest, facilitating the analysis of the influence of topology structures on the GNN model performance. (e) The Feature Matrix View visualizes the feature distribution of nodes. 
} 
\label{Fig.system}
\end{figure*}


%
%
%
%

\IEEEpeerreviewmaketitle

\IEEEraisesectionheading{\section{Introduction}\label{sec:introduction}}

\IEEEPARstart{G}{raphs} are pervasive in various applications, such as citation networks, social media, and biology.
Analyzing graph data helps us understand the hidden patterns in graphs and benefits many graph-related tasks, including node classification, link prediction, and graph classification.
For example, an effective analysis of a paper citation graph can facilitate the prediction of a new paper~\cite{kipf2016semi,hamilton2017inductive}.
An exploration of social networks can benefit the creation of an adaptive friend recommendation system in social media~\cite{DBLP:journals/access/ChenXZZX20}.
By modeling molecules as graphs, where atoms and chemical bonds are treated as nodes and edges respectively, we can build machine learning techniques to predict the chemical properties (\eg, solubility) of chemical compounds~\cite{DBLP:conf/nips/FoutBSB17}.

In recent years, graph analytics has embraced a new breakthrough---Graph Neural Networks (GNNs).
A fast growing number of GNN models have been proposed to solve graph-based tasks.
For example, Graph Convolutional Network (GCN)~\cite{kipf2016semi} adapts the convolutional operation from natural images to graphs and conducts semi-supervised learning to perform node classification on them.
Graph Attention Network (GAT)~\cite{velivckovic2017graph} further integrates the attention mechanism, which is widely used in Natural Language Processing (NLP), into the GNN model architecture and dynamically assigns weights to different neighbors to enhance the model performance.
The advances of GNNs bring new opportunities to the analysis of graph data and have become increasingly popular in recent years.
However, similar to other deep neural networks, GNN models also suffer from the difficulty of interpreting their working mechanisms.
When developing or using GNNs, developers and users often need to evaluate the model performance and explore the causes of model errors and failures, which, unfortunately, is often hard to achieve. 
Therefore, how to enable convenient error diagnosis of GNN models has become a challenging but significantly important task.


Visualization has been applied to helping model developers devise new deep learning techniques, and debug and compare different types of deep neural networks~\cite{hohman2018visual}.
For example,
various visualization techniques have been proposed to facilitate the development of a variety of deep learning models, such as CNN \cite{liu2016towards}, RNN \cite{ming2017understanding}, GAN \cite{wang2018ganviz}, and DQN \cite{wang2018dqnviz}. 
These visualizations have achieved great success in understanding and analyzing those deep learning models.
However, it is very challenging to directly apply them to GNNs, since most of those techniques are exclusively designed for Euclidean data like images and text, while GNNs mainly work on non-Euclidean data such as graphs.

Another challenge for the error diagnosis of GNNs comes from the fact that GNNs often involve both the complex topological structures and high dimensional features of graphs, as well as the interplay between them.
To effectively analyze GNNs, it is crucial to properly link the topological data, high dimensional features, and prediction results with a comprehensive workflow. 
Preliminary studies~\cite{baldassarre2019explainability,ying2019gnnexplainer,li2020explain} have proposed techniques to explain GNN model prediction results.
Most of them focus on instance analysis, \ie, explaining a prediction for single nodes.
However, there still lacks the ability and research at a higher level, \ie, analyzing and understanding the common causes of the classification errors of groups of nodes.
Their methods make it difficult to conveniently explore the general error patterns in the prediction results of a GNN model, as well as further gain insights for model improvement.
In summary, it still remains unclear how to develop new visualization techniques to facilitate the effective error diagnosis of GNNs.

In this paper, we propose a novel error-pattern-driven visual analytics system, {\name}\footnote[1]{https://gnnlens.github.io/}, to provide model developers and users with deep insights into model performance and its dependency on data characteristics.
Instead of analyzing the GNN prediction results of single instances, we investigate the patterns in the prediction results shared by a group of instances to obtain generalizable insights into the model architecture.
We worked closely with two GNN experts for four months to derive the design requirements of {\name}.
{\name} comprises five views: Control Panel,  Parallel Sets View, Projection View, Graph View, and Feature Matrix View, as shown in Fig.~\ref{Fig.system}.
The Parallel Sets View enables users to see the distribution of node-level metrics (Fig.~\ref{Fig.system}(b)). 
The Projection View presents a set of 2D projections of the selected nodes according to metrics summarized from different perspectives, enabling users to extract potential clusters of nodes (Fig.~\ref{Fig.system}(c)).
\zhihua{}{Node glyphs are proposed to help users conveniently learn about multiple metrics of nodes and extract general error patterns in the Projection View.}
The Graph View shows the node information and topological structure of the whole graph. Users can also zoom in and focus on the topological structure of a specific node (Fig.~\ref{Fig.system}(d)). The Feature Matrix View shows the detailed features of selected individual nodes (Fig.~\ref{Fig.system}(e)). All the five views of {\name} are linked together to help users analyze GNN models simultaneously from multiple angles and facilitate the exploration of error patterns of the GNN models.
We conducted two case studies and expert interviews to demonstrate the effectiveness and usability of {\name} in helping model developers understand and diagnose GNNs.

The contributions of our work are listed as follows:
\begin{itemize}
  \item  A visual analytics system to assist model developers and users in understanding and diagnosing GNNs.
  \item Case studies on analyzing error patterns in GNN prediction results and interviews with domain experts to demonstrate the effectiveness and usability of the proposed system.
 
\end{itemize}

The remainder of this paper is organized as follows. Section~\ref{sec:related_work} discusses the related work of this paper, including GNNs, visual analytics in deep learning, and GNN explainability. Section~\ref{section:background} provides a brief introduction to the basic concepts of GNNs, such as typical architectures GCN and GAT.
By working closely with domain experts, we summarize the design requirements of understanding and diagnosing GNN models in
Section~\ref{subsec:design_requirements} and further introduce the technical details of
{\name} in Section~\ref{sec:gnnvis}. 
We evaluate our approach through case studies and expert interviews in Section~\ref{sec:evaluation} and discuss the possible limitations and future work of our approach in Section~\ref{sec:discussion}. Section~\ref{sec:conclusion} concludes the paper with a brief summary of the proposed method.

 
\section{Related Work}
\label{sec:related_work}
The related work of this paper can be categorized into three groups: graph neural networks, visual analytics in deep learning, and graph neural networks explainability.

\subsection{Graph Neural Networks}

GNNs have been developed to analyze graph data by extending CNNs or RNNs to the graph domain~\cite{zhou2018graph} in the past few years.
These neural networks have gained promising prediction results for analyzing graphs. 

For the GNNs derived from CNN, they can be categorized into spectral approaches and spatial approaches~\cite{zhou2018graph}. Spectral approaches define convolution on the spectral representation of graphs~\cite{bruna2013spectral, defferrard2016convolutional, kipf2016semi}. The work done by Bruna et al. \cite{bruna2013spectral} 
is the first attempt
to generalize the convolution concept from natural images to the graph domain. Defferrard et al. \cite{defferrard2016convolutional} approximated the spectral convolution as Chebyshev polynomials of the diagonal matrix of eigenvalues, resulting in further low computational cost. Kipf and Welling \cite{kipf2016semi} further simplified the Chebyshev polynomials by using the first order of polynomials and renormalization tricks, known as GCNs, which have inspired many follow-up studies.
Spatial approaches directly define convolution on spatially close neighbors~\cite{hamilton2017inductive, duvenaud2015convolutional, atwood2016diffusion, zhuang2018dual, niepert2016learning, gao2018large, monti2017geometric}. Hamilton et al.~\cite{hamilton2017inductive} proposed GraphSAGE, which uses sampling methods and aggregators defined over the neighborhood to reduce dependence on processing whole graphs. 
Their approach greatly accelerates the GNN used in large scale graphs. 
Another direction is to extend RNN to the graph domain. Prior studies have attempted to utilize the gate function in GNNs to improve its ability to propagate information across graph structure~\cite{li2015gated, tai2015improved, zayats2018conversation, liang2016semantic, DBLP:journals/tvcg/WangJWCMQ20}. 

Researchers have also made significant progress in analyzing GNN models. For example, Li et al.~\cite{li2018deeper} showed that the graph convolution of a GCN is merely a Laplacian smoothing operation
but when the number of layers increases, the risk of over smoothing will be increased.
Also, they showed that when few training labels are given to train GCN models, co-training methods and self-training methods will improve the performance of GCN models. Xu et al.~\cite{xu2018powerful} provided a theoretical framework to analyze expressive power for GNNs and proved that their proposed model is as expressive as the Weisfeiler Lehman graph isomorphism test. 
Different from these studies, this study is aimed at extracting general error patterns of GNN models and further helps model developers understand and diagnose the models.

\subsection{Visual Analytics in Deep Learning}

Nowadays, there is a growing trend to use visualizations to understand, compare, and diagnose deep neural networks~\cite{hohman2018visual}. 
Prior studies on using visual analytics to enhance the interpretability of deep neural networks can generally be categorized into two types: model-agnostic visualizations and model-specific visualizations.
For model-agnostic visualizations, prior studies mainly focus on visualizing the model input and output to provide insights into the correlation between them~\cite{Zhang2018manifold, alsallakh2014visual} or using surrogate models to explain the deep neural networks~\cite{DBLP:journals/tvcg/MingQB19,DBLP:journals/tvcg/WangGZYS19}.
However, these model-agnostic visualizations avoid showing the hidden states of the deep neural networks and fail to reveal the inner working mechanism of different models.

To support a dive into the deep learning models, researchers have also proposed a series of model-specific visualizations for explaining deep learning models.
Previous model-specific visualizations have covered a wide range of deep learning models, including CNN, RNN, and GAN.
A variety of visualization techniques and interactions have been designed based on the data type, the model structures, and the working mechanism of different deep learning models.
Since CNN and RNN are the most widely-used deep learning models~\cite{lecun2015deep, DLBook}, a majority of model-specific visual analytics are proposed for both types of models. 
For example,
CNNs are usually modeled using the directed acyclic graph visualization, and the output of each layer is usually displayed using matrix-based visualizations~\cite{liu2016towards, liu2018deeptracker, pezzotti2017deepeyes}.
To open the black box of RNNs, clustering methods and correlation visualizations have been proposed to uncover the dynamic hidden states and learned patterns in RNNs \cite{ming2017understanding, strobelt2017lstmvis, strobelt2018s}. 
Recently, visual analytics methods tailored for generative models~\cite{wang2018ganviz, liu2017analyzing, kahng2018gan} 
and reinforcement learning models~\cite{wang2018dqnviz} have also been proposed.

Despite much work having been done by using visualization approaches to improve the explainability of deep learning models, little research has been conducted to enhance the explainability of GNNs through visualizations.  The most similar work is called CorGIE~\cite{liu2021visualizing}, which considers the correspondence between graph topology, node features, and latent embedding to understand GNNs. 
Different from CorGIE, this paper focuses on analyzing error patterns of GNNs and proposes a visual analytic approach to facilitate the error diagnosis of GNNs.
It mainly enables users to inspect various kinds of node metrics reflecting the influence of training nodes and compare prediction results of multiple models to analyze error patterns of the GNNs.

\subsection{Graph Neural Network Explainability}


According to our survey, only a few  studies have attempted to explain GNN models.
For instance,
Baldassarre et al. \cite{baldassarre2019explainability} explored the possibilities of adapting explanation techniques from CNNs to GNNs.
They empirically evaluate three widely-used CNN explanation methods, \ie, Sensitivity Analysis (SA), Guided Back Propagation (GBP), and Layer-wise Relevance Propagation (LRP) when explaining GNN decisions. 
They found that explanations produced by SA or GBP tend to
be inconsistent with the human interpretation,
while LRP produces more natural explanation results.
Meanwhile,
Ying et al.~\cite{ying2019gnnexplainer} proposed GNNExplainer, which uses a subgraph to explain the GNN model prediction. 
Given a trained GNN model, they formulate an optimization task to maximize the mutual information between the trained model prediction and distribution of possible graph structures. They regard the subgraph as the explanation results. Li et al. \cite{li2020explain} further extended GNNExplainer which is designed for an undirected unweighted graph to suit a directed weighted graph.  

\mzhihua{}{Previous studies have mainly focused on providing an instance-based explanation of the prediction results of GNN models. Different from them, our work mainly focuses on analyzing error patterns made by GNN models, including group-level and instance-level errors.}

\begin{figure}[htb]
\centering 
\includegraphics[width=0.48\textwidth]{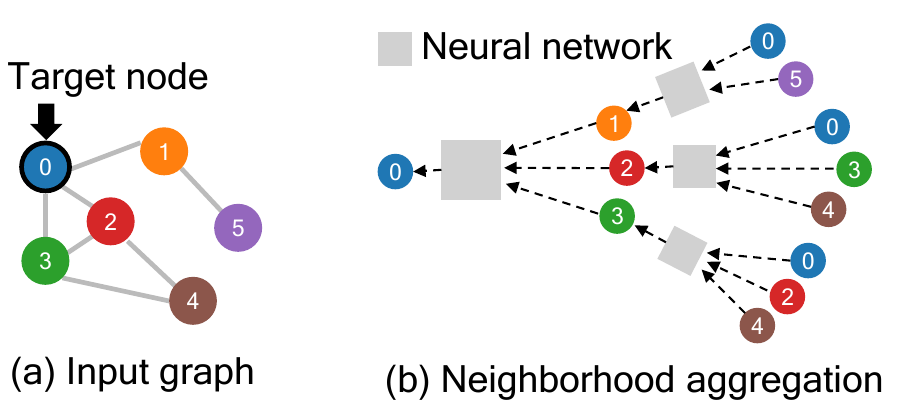}
\caption{
Given an input graph (a), GNN predicts the label of the target node (\eg, the blue node) by aggregating the information from neighboring nodes (b).
} 
\label{Fig.gnn}
\end{figure}

\section{Background}
\label{section:background}


GNNs
are deep neural networks that directly operate on graphs (\ie, networks).
A graph can be represented as $G=(V,E)$, where $V$ denotes the vertex set and $E$ denotes the edge set.
$X \in \mathbb{R}^{N\times d}$ is the feature matrix of the graph, where $N$ denotes the number of nodes in the vertex set and $d$ is the dimension of each node feature. 
The labels of the nodes in the graph are often denoted as
$Y$.
In this paper, we do not consider the features in edges.

We adopt similar notations introduced in~\cite{fey2019fast} to illustrate the concept of GNNs. GNNs can be generally expressed in a neighborhood aggregation or message passing scheme~\cite{hamilton2017representation}, as shown in Fig.~\ref{Fig.gnn}. A general message passing function for GNN is shown as below:

\begin{equation}\mathbf{x}_{i}^{(k)}=\mathcal{C}^{(k)}\left(\mathbf{x}_{i}^{(k-1)}, \mathcal{A}_{j \in \mathcal{N}(i)} \phi^{(k)}\left(\mathbf{x}_{i}^{(k-1)}, \mathbf{x}_{j}^{(k-1)}\right)\right)\end{equation}

where $x_{i}^{(k-1)}$ denotes node features of Node $i$ in Layer $k-1$. 
$\mathcal{N}(i)$ denotes the neighborhood of Node $i$. $\mathcal{A}$ denotes a differentiable permutation-invariant function, e.g., sum. $\mathcal{C}$ and $\phi$ denote differentiable functions such as Multi-Layer Perceptrons (MLPs). 

GCNs and GATs are two popular
GNNs. 
According to the study by  Kipf et al.~\cite{kipf2016semi}, the message passing function of GCN can be defined as follows:

\begin{equation}
\mathbf{x}_{i}^{(k)}=\sigma \left( \sum_{j \in \mathcal{N}(i) \cup\{i\}} \frac{1}{\sqrt{\operatorname{deg}(i)}  \sqrt{\operatorname{deg}(j)}} \left(\mathbf{W} \mathbf{x}_{j}^{(k-1)}\right) \right)\end{equation}

where the features of the neighbors of Node $i$ are first transformed by a weight matrix $\mathbf{W}$. Then they are normalized by their degree and finally summed up.
\zhihua{}{Finally, a non-linear activation function $\sigma$ is applied to process it.}

GAT is first proposed by Veli{\v{c}}kovi{\'c} et al.~\cite{velivckovic2017graph} and its message passing function is defined as follows:
\begin{equation}\mathbf{x}_{i}^{(k)}=\sigma\left(\sum_{j \in \mathcal{N}(i)\cup\{i\}} \alpha_{i j} \mathbf{W} \mathbf{x}_{i}^{(k-1)}\right)\end{equation}

where $\mathbf{W}$ and $\sigma$ are similarly defined as above. 
Different from GCN, GAT assigns different weights (attention coefficients) to each neighbor. The attention coefficients $\alpha_{i, j}$ are computed as:
\begin{equation}
\alpha_{i, j}=\frac{\exp \left(\text { LeakyReLU }\left(\mathbf{a}^{\top}\left[\mathbf{W} \mathbf{x}_{i} \| \mathbf{W} \mathbf{x}_{j}\right]\right)\right)}{\sum_{k \in \mathcal{N}(i) \cup\{i\}} \exp \left(\text { LeakyReLU }\left(\mathbf{a}^{\top}\left[\mathbf{W} \mathbf{x}_{i} \| \mathbf{W} \mathbf{x}_{k}\right]\right)\right)}
\end{equation}
where $\mathbf{a}$ is a weight vector and LeakyReLU is an activation function which is defined as $\text{LeakyReLU}(x)=\max (0, x)+\text{negative\_slope}*\min(0,x)$. 

The GNN models are mainly applied to the tasks of node classification and link prediction in individual graphs.
In this paper, we take node classification as an example to illustrate how our approach can improve the interpretation of GNN models and facilitate model diagnosis.
Such kinds of node classification tasks are often done in a semi-supervised way. 
Given a set of labeled nodes (\ie, training nodes) in a graph, 
a GNN model will be trained to predict the labels of the rest of the nodes in the graph.


\section{Design Requirement Analysis}
\label{subsec:design_requirements}
We work closely with two GNN experts, who are also co-authors of this work, to collect their feedback on
the GNN interpretation issues they are facing and their current practices of understanding and diagnosing GNN models. 
One expert (E1) is a senior researcher who specializes in developing new kinds of GNNs. The other expert (E2) is a deep learning developer with strong experience in applying GNNs to modeling and analyzing the topology data from different application domains such as online education and visualization.
Also,
the development of {\name} was conducted in an iterated way. After we finished each version of the system, we asked experts to use the pilot system, comment on the limitations, and suggest possible improvements.
By combining the original requirements proposed by the experts and their subsequent comments on the limitations of the systems, we complied a list of major design requirements proposed by the domain experts, which can be summarized as follows:

\textbf{R1: Provide an overview of GNN results.} 
All experts commented that an overview of the GNN performance is crucial for the GNN analysis.
To gain an overview of the dataset and classification results, the system needs to summarize various types of information, such as degree distribution and ground truth label distribution. 
This information, covering various aspects of a GNN model, needs to be organized and presented in a clear manner.
Meanwhile, the correlation among this information should be presented to help users develop initial hypotheses about any possible error patterns in GNN results, \ie, a set of wrong predictions that share similar characteristics.

\textbf{R2: Identify error patterns.}
After developing initial hypotheses about the error patterns, users need more detailed information to verify them. 
Specifically, users need to examine the characteristics shared by a set of wrong predictions and verify whether error patterns formed by these characteristics make sense in analyzing GNNs based on their domain knowledge.
During the interview, experts agreed that they usually use several characteristics to group the wrong predictions and identify error patterns.
For example, one expert stated that \textit{``misclassified nodes usually have a relatively large shortest path distance to the labeled nodes.''}
Therefore, the system should support users in examining these characteristics and identifying error patterns.

\textbf{R3: Analyze the cause of error patterns.}
After identifying error patterns, finding the causes of these errors is important for users to understand, diagnose, and improve the GNNs.
More detailed information is needed to understand the possible causes of error patterns.
Specifically, users need to inspect the graph structures and node features to determine the causes of error patterns.
According to the feedback from expert interviews, there are two main sources of wrong GNN predictions: noise in the training data and inaccurate feature aggregation in GNNs.
To predict the label of a node, GNN aggregates the node's own feature with the features of the neighboring nodes at each layer.
Noise in the training data, e.g., the same nodes but different labels, can confuse the GNN and lead to wrong predictions.
Inaccurate feature aggregation at any layer will also influence the GNN prediction of the node.



\section{GNNLens}
\label{sec:gnnvis}

This section describes the details of the proposed approach, {\name}. We first provide a system overview. Next, detailed information on proxy models and metrics is provided. Finally, we introduce the detailed visualization design of each view in {\name}.


\subsection{System Overview}

{\name} consists of three major modules: storage, data processing, and visualization. 
The storage module stores and manages graph data and models. The data processing module implements the necessary procedures for analyzing the graph and model predictions, especially for calculating various kinds of metrics. 
The processed data is then passed to the visualization module, which supports the interactive visual analysis of the GNNs.
\zhihua{}{The storage and data processing modules are developed using Python and Deep Graph Library (DGL)~\cite{wang2019dgl} and integrated into a back-end web server that is built with Flask.}
The GNN models are implemented with PyTorch.
We implement the visualization module as a front-end application using React, Typescript, and D3.

\subsection{Proxy Models Training and Metrics Definition}
\label{sec:models_metrics}

Inspired by the fact that GNN prediction results are influenced by both graph structure and node features~\cite{zhou2018graph}, we define two proxy models to analyze the influence of the graph structure and node features on GNN prediction results.
Through expert interviews, experts are concerned about whether the graph structure or node features have a greater impact on GNN prediction, and then determine which components will have more impact. 
Hence, similar to the ablation study when evaluating GNN models~\cite{xu2018powerful}, we define two proxy models such as GNN Without Using Features (GNNWUF) and Multi-Layer Perceptron (MLP).
The two models are chosen, since the two proxy models have the same model architectures as the GNN but are trained using different input data. GNNWUF is trained only using the graph structure while MLP is trained only using the node features.
When training GNNWUF, we use one hot encoding as the node feature for each node, meaning GNNWUF considers only the graph structures.
When GNN considers only the features of the node itself, then it can degenerate into an MLP model. 
Hence, MLP is chosen as the other GNN proxy model that only considers the node features and is used to evaluate the influence of node structures.
We train both proxy models with the same settings as the training of GNN.

To further help users understand the impact of the graph structure and node features, we also provide a number of metrics, including graph structure based metrics that take into account the graph structure but ignore the node features, and node feature based metrics that take the node features into account but ignore the graph structure.
Those metrics are derived from expert interviews.
Details are presented in the following paragraphs.



    
\begin{figure}[htb]
\centering 
\includegraphics[width=\linewidth]{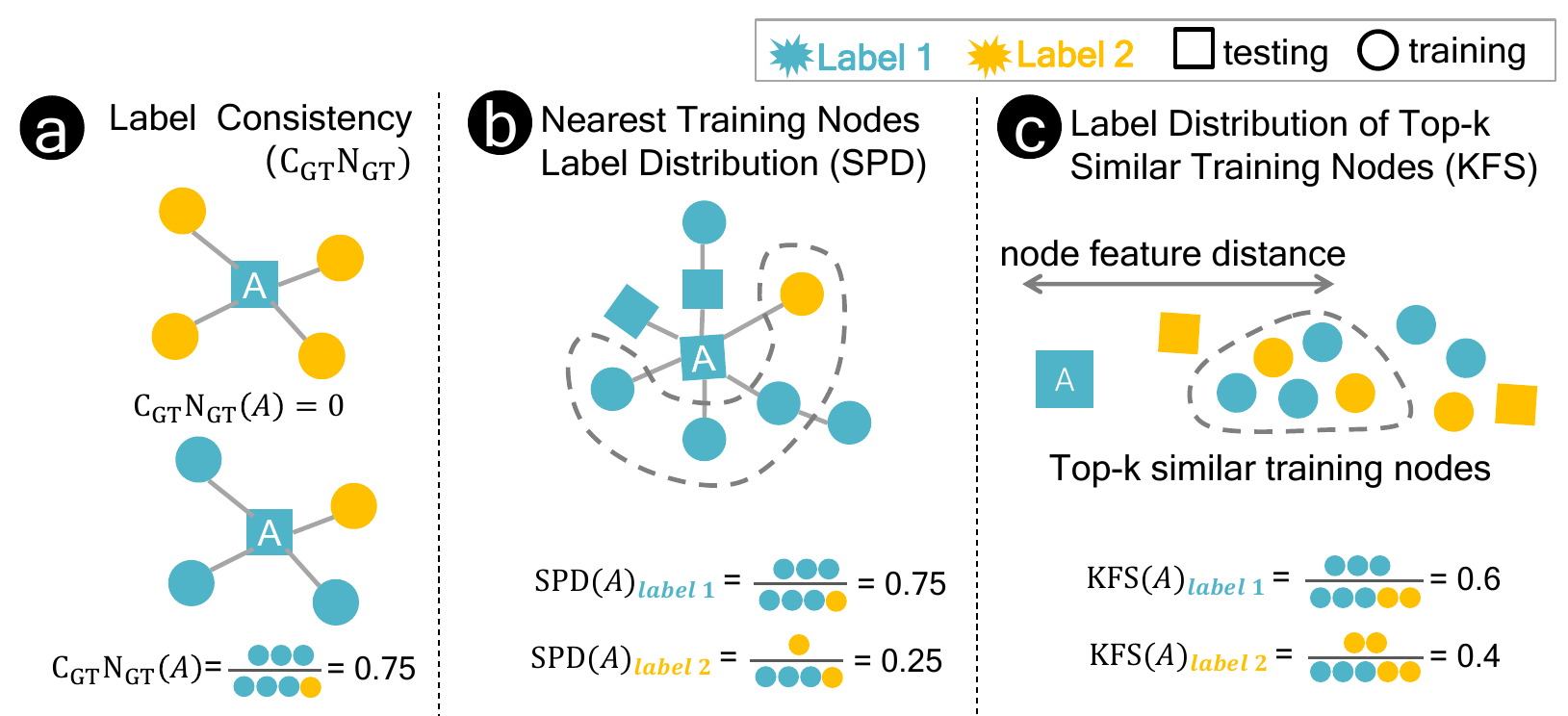}
\caption{Examples illustrating the computation of metrics, including Label Consistency (a), Nearest Training Nodes Label Distribution (b), Label Distribution of Top-k Similar Training Nodes (c). } 
\label{Fig.gnnvis_metrics}
\end{figure}

\subsubsection{Graph Structure based Metrics}
\begin{itemize}
    
    \item \emph{Node degree}. GNNs mainly operate over the neighborhood of each node and the number of neighbors can affect the final performance of a GNN model~\cite{demonet_kdd19}.
    Therefore, the \textit{node degree} is considered in this study. 
     \item \emph{Center-neighbor consistency rate}. The \textit{center-neighbor consistency rate} depicts how consistent the \textit{labels} of the \textit{current node} and its surrounding neighbors are~\cite{xu2020label}.
    It can be divided into four major categories by considering both the \textit{ground truth labels} and their predictions:
    (1) \textbf{\textit{Label consistency}} shows the percentage of neighbors that have the same \textit{ground truth label} as the \textit{current node} (Fig.~\ref{Fig.gnnvis_metrics}(a)); 
    (2) \textbf{\textit{Label-Prediction consistency}} describes the percentage of neighbors whose \textit{GNN prediction labels} are the same as the \textit{current node's ground truth label}; 
    (3) \textbf{\textit{Prediction-Label consistency}} delineates the percentage of neighbors whose \textit{ground truth labels} are the same as the \textit{current node's GNN prediction label}; 
    and (4) \textbf{\textit{Prediction consistency}} refers to the percentage of neighbors which have the same \textit{GNN prediction label} as the \textit{current node}.
    These values can indirectly reflect how many neighbors satisfy the constraints.
    If the \textit{node degree} is zero, then the \textit{consistency rate} is set to zero.
    These metrics help users check whether the one-hop neighborhood exerts influence on the \textit{GNN prediction result} on the node of user interest.
    \item \emph{Shortest path distance to training nodes}. We use the breadth-first search (BFS) algorithm to calculate the \textit{shortest path distance from the current node to the training nodes}.  
    The algorithm will first start traversing the \textit{current node} and then the neighbors of the visited nodes. When it first detects a node in the training set, the algorithm will regard the distance from that node to the \textit{current node} as the \textit{shortest path distance from the current node to the training nodes}.
    The distribution of the training nodes, also called labeled nodes, can have a significant influence on \textit{GNN prediction}~\cite{yang2019spagan}.
    \item \emph{Nearest training nodes label distribution}. 
    To investigate the influence of training nodes distribution on model training, we calculate the \textit{nearest training nodes label distribution}.
    To calculate the \textit{label distribution of the nearest training node(s)}, we first find the closest training nodes
    to the \textit{current node} in terms of shortest path distance.
    Then we count the frequency of the labels of these training nodes and normalize them into $[0,1]$. The normalized frequencies are considered to be the \textit{nearest training nodes label distribution} (Fig.~\ref{Fig.gnnvis_metrics}(b)). Analyzing these metrics helps to investigate the influence of training nodes distribution on model training~\cite{yang2019spagan}.

    \item \emph{Nearest training nodes dominant label consistency.}
    In order to help users quickly capture the dominant information of \textit{nearest training nodes label distribution} and further diagnose the causes of errors in \textit{GNN prediction results}, we define the \textit{nearest label} as the label that most frequently occurs in the training nodes closest to a specific node in terms of topological distance.
    Then, we further consider whether the \textit{nearest label} is consistent with the \textit{ground truth label} of this specific node. If yes, we set the \textit{nearest training nodes dominant label consistency} for this node to \textit{True}; otherwise, it is set to \textit{False}. Sometimes, there may be multiple \textit{nearest labels}. Then we directly set the \textit{nearest training nodes dominant label consistency} to \textit{Not Sure}.
    Such a metric is derived from \textit{nearest training nodes label distribution} and the \textit{ground truth label} of the \textit{current node}. 
    If it is true, the current node can get information from the structure and the training nodes and it has a high chance of being correctly classified. Otherwise, it has high probability to be misclassified~\cite{yang2019spagan}.
\end{itemize}
\subsubsection{Node Features based Metrics}
\begin{itemize}
    \item \emph{The label distribution of the top-k training nodes with the most similar features.}
    The feature similarity between two nodes is defined as the cosine distance between the feature vectors of two nodes~\cite{chen2020iterative}. We first find the \textit{top-k training nodes with the most similar features} to the node of user interest. Then we count the frequency of labels of those training nodes and normalize them into $[0,1]$. They are
    defined as
    the \textit{label distribution of the top-k training nodes with the most similar features} (Fig.~\ref{Fig.gnnvis_metrics}(c)).
    With this metric, we can analyze the influence of node features on \textit{GNN predictions}. We set the default value of k to 5 in our current implementations.
    
    \item \emph{Top-k most similar training nodes dominant label consistency.}
   Similar to the definition of the previous metric, we can also calculate the \textit{top-k most similar training nodes dominant label consistency} in the same way. The major difference is that this metric reflects the influence of training node features on the model prediction results on the current node~\cite{chen2020iterative}.
\end{itemize}

\subsection{Visualization}

As shown in Fig.~\ref{Fig.system}, {\name} visualization module consists of a Control Panel (a), a Parallel Sets View (b), a Projection View (c), a Graph View (d), and a Feature Matrix View (e).
\mzhihua{}{The Control Panel allows users to choose a graph dataset, select inspected models (Appendix A),
and explore different subsets of the dataset (\eg, \textit{all, training, validation,} and \textit{testing}).} It also allows users to choose the k value to calculate node features based metrics. 
The Parallel Sets View (Fig.~\ref{Fig.system}(b))
visualizes the distribution of node-level metrics, which are defined in Section~\ref{sec:models_metrics}. Users can select a subset of metrics to inspect their distribution and correlation. Then users can select a subgroup of nodes and check them in the Projection View (Fig.~\ref{Fig.system}(c)).
The Projection View presents a set of 2D projections of the selected nodes according to metrics summarized from different perspectives. Users can lasso-select a cluster of nodes in one plane to see their locations on the whole graph in the Graph View (Fig.~\ref{Fig.system}(d)) and their feature distributions in the Feature Matrix View (Fig.~\ref{Fig.system}(e)).

\subsubsection{Parallel Sets View}
In order to provide a high-level summary (\textbf{R1}) and help users understand the datasets and identify error patterns of \textit{GNN models prediction} (\textbf{R2}), we design the Parallel Sets View to visualize node-level metrics using Parallel Sets~\cite{bendix2005parallel}.
Previous work \cite{ren2016squares, DBLP:journals/tvcg/WexlerPBWVW20, pezzotti2017deepeyes, DBLP:journals/tvcg/DingenVHMKBW19, DBLP:journals/tvcg/KahngAKC18} explored the selection of a subset of sample properties to study machine learning models. Inspired by previous research, we use this strategy to investigate error patterns in GNN models. We propose to use Parallel Sets to investigate error patterns in GNN prediction results, following the previous work \cite{vosough2018using, DBLP:conf/vda/Chaudhuri18}.
 Users can select what metrics are to be displayed in the Parallel Sets through Parallel Sets Settings Modal. In general, displaying fewer than five axes in Parallel Sets is a good practice to reduce visual clutter and make efficient use of functions in Parallel Sets. Due to the constraint that the Parallel Sets are used to display the categorical variables, we need to convert the continuous metrics to categorical variables by grouping a range of values into one category. Then we can also show them in the Parallel Sets View.

As shown in Fig.~\ref{Fig.system}(b), 
each axis of the Parallel Sets shows a categorical variable. The axis is partitioned into multiple segments representing different categories of the variable. The width of each segment represents the number of nodes falling into that category. We can directly see the distribution of the categories on the axis. Between two consecutive axes, multiple ribbons are shown to connect the two axes, each simultaneously representing the nodes that satisfy the conditions specified by the two axes.

Users can easily select a subset of nodes in the dataset and further investigate their node metrics and the \textit{GNN model prediction results}.
When users click on a segment, the corresponding category of that axis will be selected. Also, when users click on the ribbon in the Parallel Sets, the corresponding set of nodes will be selected.
Besides, the axes in the Parallel Sets can be easily reordered by users through drag-and-drop.
By filtering the nodes according to node-level metrics such as \textit{correctness} and \textit{label}, users can easily select a node subset of their interest for further analysis.  

A common alternative for visualizing multivariate data is the Parallel Coordinates Plot (PCP)~\cite{inselberg1990parallel}.
Each data point is visualized as a single line across different attributes.
However, when it comes to categorical data, it is challenging to identify the proportions of data that fall into specific categories. Compared with PCP, Parallel Sets intuitively show the distribution of the categories in each axis and the correlation between multiple axes. Thus, 
Parallel Sets are chosen to display the overall distribution of node attributes.


\begin{figure*}[htb]
\centering 
\includegraphics[width=\textwidth]{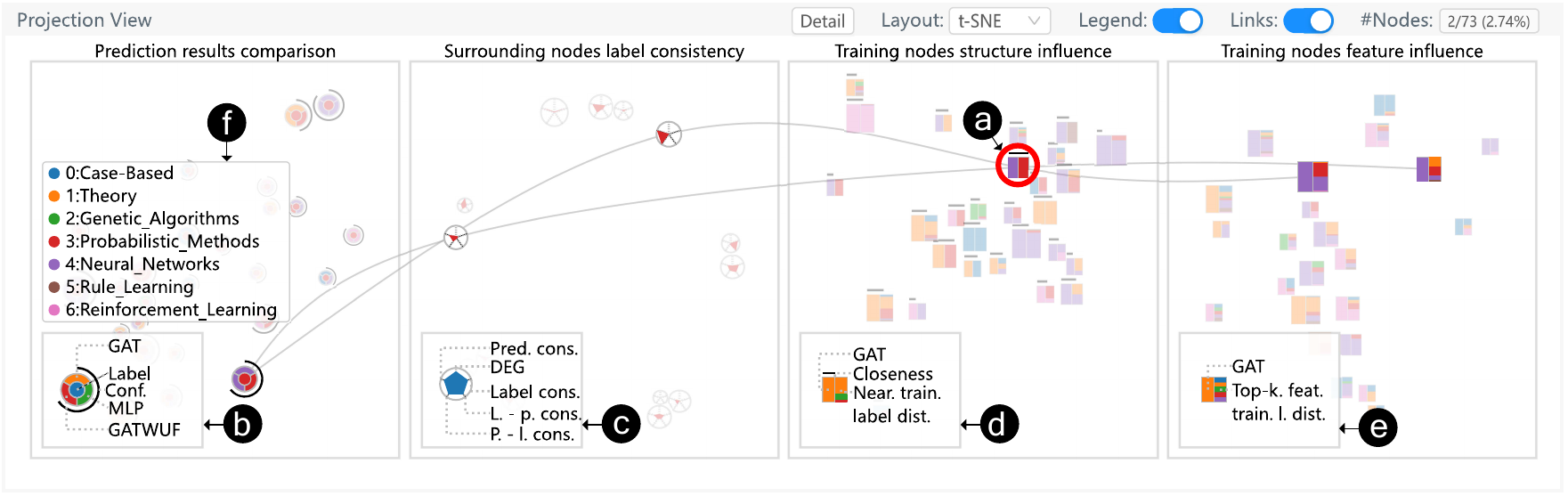}
\caption{\zhihua{}{Projection View enables users to analyze the node-level metrics of the subset of nodes from different perspectives: (a) The links connecting different kinds of information of the same nodes shown in different planes will be displayed when users lasso a group of nodes in one plane; (b-e) Node glyphs design in planes of the Projection View; (f) Color indicates the corresponding label.}} 
\label{Fig.projection_view_overall}
\end{figure*}

\subsubsection{Projection View}
\label{sec:ProjectionView}

With the overview of the dataset and GNN models provided by the Parallel Set View, we further design the Projection View to give users more insights into the subset of nodes selected in the Parallel Sets View (\textbf{R2, R3}). We group a subset of node-level metrics, display them in glyphs, and further project them to the 2D plane.
The Projection View allows users to investigate the similarity of nodes regarding different perspectives.
It can be helpful for investigating whether the nodes with similar node metrics share similar error patterns. 

In the Projection View, we provide a set of linked projection planes of the nodes that use different features. 
Different from similar designs in EmbeddingVis~\cite{DBLP:conf/ieeevast/LiNHCYM18}, we design different node glyphs to display different combinations of node-level metrics.
To project those node glyphs in 2D plane, users can choose to use the t-SNE~\cite{maaten2008visualizing} or UMAP~\cite{mcinnes2018umap} projection as the basic layout algorithm. Moreover, the force-directed collision-avoidance method is integrated into the basic layout algorithm to prevent the overlapping of node glyphs.
When users lasso-select a set of nodes in a projection plane, the links between the same nodes in different planes will be shown to help users identify the nodes and other aspects of those nodes' properties, as shown in Fig.~\ref{Fig.projection_view_overall}. After users hover on the node glyphs, the legend and detailed information of those node glyphs will be displayed.
However, due to the limited screen space, it cannot display hundreds let alone thousands of node glyphs. 
\zhihua{}{Therefore, we apply a hierarchical clustering algorithm with complete linkage to cluster these nodes based on the corresponding distance function~\cite{clusterAnalysis}. Two clusters will be merged into one cluster when the distance of two clusters is less than or equal to a threshold, which is empirically set to 0.5. 
In terms of categorical metrics such as \textit{ground truth label} and \textit{prediction results}, the majority of a categorical metric in one cluster is regarded as the aggregate result of that metric. For continuous metrics such as \textit{center-neighbor consistency rate} and \textit{degree}, we take their average as the aggregated results.
Cluster-level node glyphs are designed based on the aggregate results of the node-level metrics for individual nodes in the clusters.}
In order to help users further inspect individual nodes, after users select a subset of cluster-level node glyphs, users can switch to "Detail" mode, then the Projection View will display individual node glyphs for nodes in such a cluster. This design greatly enhances the scalability of the Projection View.

In our implementation, we categorize the metrics into four groups and provide four projections for each group of node metrics. The four projection planes are \textit{prediction results comparison}, \textit{surrounding nodes label consistency}, \textit{training nodes structure influence}, and \textit{training nodes feature influence}, respectively. Different glyph designs are also proposed for the nodes. We will introduce them one by one in the following paragraphs.

\textbf{\textit{A. Prediction results comparison.}} 
This plane aims to help users compare different \textit{models prediction results} and reveal the relative influence of graph structures and features for each node or cluster.
The metrics used in this plane include the \textit{ground truth label} $GT$, the \textit{prediction label of three models}, \ie, GNN, GNNWUF, MLP $P=[P_1, P_2, P_3]$, and the \textit{Confidence} of GNN prediction $CONF$. 
As shown in Fig.~\ref{Fig.projection_view_overall}(b), \textit{three model prediction results} can be found in the pie chart. The inner circle encodes the \textit{ground truth label}. The outer circular ring encodes the \textit{confidence}. The radius of the whole node glyph encodes the size of the clusters. \zhihua{}{Through such a node glyph, users can easily compare the \textit{ground truth label} and \textit{model prediction results} and understand how confidently GNN models make predictions. For example, if the glyph shows consistent color across the inner circle and the three sectors of the pie chart, it means that the model can use either the node features or the graph structure information to correctly predict the node labels.}
Through the projection, the nodes with similar metrics will be in close proximity. Users can see if there are clusters of nodes with the same \textit{ground truth labels} and \textit{predictions}, which helps GNN model developers and users further analyze what causes the model to make such predictions. 
For projection and clustering, the distance between Node $a$ and Node $b$ in this plane is defined as below:

\mzhihua{}{
\begin{equation}
\label{eq:plane_a}
\begin{split}
    D_{1}^{2}(a,b) = \mathbb{I}\{GT_a\ne GT_b\}+\sum_{j=1}^{3}\mathbb{I}\{P_{aj}\ne P_{bj}\}\\+(CONF_{a}-CONF_{b})^{2}
\end{split}
\end{equation}
}

\mzhihua{}{where $\mathbb{I}\{*\}$ is an indicator function, which will be 1 if the expression is true and 0 otherwise.}
Such a distance function guarantees that the value of each term is between 0 and 1.

\textbf{\textit{B. Surrounding nodes label consistency. }}
To help users explore the \textit{label consistency} between a node and its neighboring nodes, 
we show the \textit{ground truth label} $GT$, the \textit{degree} $DEG$, the \textit{center-neighbor consistent rate} $CN = [C_{GT}N_{GT}, C_{GT}N_{PT}, C_{PT}N_{GT}, C_{PT}N_{PT}] \in [0,1]^{4}$ in this plane, where $C_{GT}N_{GT}$ represents \textit{Label consistency}, $C_{GT}N_{PT}$ represents \textit{Label - Prediction consistency, $C_{PT}N_{GT}$ represents \textit{Prediction - Label consistency}, and $C_{PT}N_{PT}$ represents \textit{Prediction consistency}.}
The node glyph (Fig.~\ref{Fig.projection_view_overall}(c)) is designed to show this group of metrics. The design is inherent from the Radar Chart as it can display continuous variables. The color of polygon encodes the \textit{ground truth label}. \zhihua{}{
When the five axes of the Radar Chart reach a large value,
it means that the \textit{center node} has many surrounding nodes, and the \textit{ground truth labels} and \textit{GNN predictions results} of that node are very consistent with the surrounding nodes. 
It suggests that the GNN model can effectively use the information of surrounding nodes for correct classification.} The radius of the whole node glyph encodes the size of the clusters. Clusters may appear and users can easily spot them, since there will be a certain shape among those node glyphs. For the projection and clustering, the distance between Node $a$ and Node $b$ is defined as:


\mzhihua{}{
\begin{equation}
\label{eq:plane_b}
\begin{split}
D_{2}^{2}(a,b) = (Norm(DEG_{a})-Norm(DEG_{b}))^{2} \\ + \mathbb{I}\{GT_a\ne GT_b\} +\sum_{i=1}^{4}(CN_{ai}-CN_{bi})^2,
\end{split}
\end{equation}}

where $Norm(d)\in[0,1]$ is the normalized degree.

\textbf{\textit{C. Training nodes structure influence.}}
To help users capture the structure influence of training nodes on \textit{GNN model prediction}, 
the metrics visualized in this plane include
\textit{GNN prediction label} $P_1$, \textit{shortest path distance to training nodes} $DIS$, and \textit{normalized nearest training nodes label distribution} $SPD\in [0,1]^{C}$. Here $C$ is the number of classes. In order to encode the $DIS$ in the node glyph and highlight the difference between smaller values, \ie, $DIS\in[0,5)$, 
we define  $Closeness=max(0,1-DIS*0.2)$. It depicts the \textit{closeness of the nearest training nodes to the current node}.
The node glyph (Fig.~\ref{Fig.projection_view_overall}(d)) is designed to show this group of metrics. 
The length of the line on the top of the rectangle encodes the \textit{closeness}. The rectangle on the right-hand side of the glyph shows the \textit{distribution of ground truth labels of training nodes with the shortest path distance to that node}. The width and height of the whole node glyph encode the size of the clusters. 
The left-hand side rectangle encodes the \textit{GNN prediction label}. This helps users analyze the correlation between those variables. \zhihua{}{When the color of the left-hand side rectangle is consistent with that of the right-hand side rectangle, it means that there is a high correlation between $P_1$ and the dominant component of $SPD$, indicating that the closest training nodes have a strong influence on \textit{GNN's prediction result of the current node}.} For the projection and clustering, the distance between Node $a$ and Node $b$ is defined as:

\mzhihua{}{
\begin{equation}
\label{eq:plane_c}
\begin{split}
  D_{3}^{2}(a,b) = \mathbb{I}\{P_{a1}\ne P_{b1}\} + +\sum_{i=1}^{C}(SPD_{ai}-SPD_{bi})^2 \\+ (Closeness_a-Closeness_b)^2.  
\end{split}
\end{equation}
}


\textbf{\textit{D. Training nodes feature influence.}} 
We further use another plane to help users capture the feature influence of training nodes. 
The metrics we used in this plane include \textit{GNN prediction label} $P_1$ and $KFS\in [0,1]^{C}$, the \textit{label distribution of the top-k training nodes with the most similar features}. 
The node glyph (Fig.~\ref{Fig.projection_view_overall}(e)) shares a similar visual design with Fig.~\ref{Fig.projection_view_overall}(d). 
The difference is that the right-hand side rectangle encoded the \textit{top-k feature similarity training nodes ground truth label distribution} and the node glyphs do not have a line at the top. 
It enables users to analyze \textit{GNN prediction results} from the perspective of features. \zhihua{}{If users find that the color of the GNN prediction result occupies a large portion of the right-hand side rectangle, it indicates that the \textit{top-k feature similarity training nodes} have a strong influence on \textit{GNN's prediction of the current node}.}
The clusters on the projection plane indicate nodes that have been similarly affected by the features. Combined with the Feature Matrix View, we can determine which feature may play a better or worse role in the GNN predictions. We use a similar distance function defined in the plane of \textit{training nodes structure influence}:

\mzhihua{}{
\begin{equation}
\label{eq:plane_d}
D_{4}^{2}(a,b) = \mathbb{I}\{P_{a1}\ne P_{b1}\} + \sum_{i=1}^{C}(KFS_{ai}-KFS_{bi})^2.
\end{equation}}

\mzhihua{}{
\textit{Weights of terms in the above distance functions:} 
For each distance function above (i.e., Equations \ref{eq:plane_a} to \ref{eq:plane_d}), we normalize each term of the function to $[0,1]$. Also, we assume each term has equal importance in differentiating the difference between nodes and empirically set the weight of each term to 1.
}

\textit{Alternative designs:} 
There are a few design alternatives for those node glyphs. For the node glyph in the plane of \textit{prediction results comparison}, we can use a $2\times2$ grid to represent the \textit{ground truth label} and \textit{three model prediction results}. However, such a design cannot effectively help users compare the metrics in the diagonal and will confuse users. Therefore, such a design is not adopted. For the node glyph in the plane of \textit{surrounding nodes label consistency}, an alternative design is to use Parallel Coordinates Plot to display the five continuous metrics. However, it is generally hard for users to distinguish between two node glyphs. For the node glyph in the plane of \textit{training nodes structure influence and feature influence}, we can use a similar node glyph in the plane of \textit{prediction results comparison} to encode the GNN prediction result in the inner circle and encode the \textit{label distribution} in the outer ring. However, to avoid any confusion in the node glyph between those planes, we do not use this design in \textit{training nodes structure influence and feature influence} plane.

\begin{figure}[htb]
\centering 
\includegraphics[width=\linewidth]{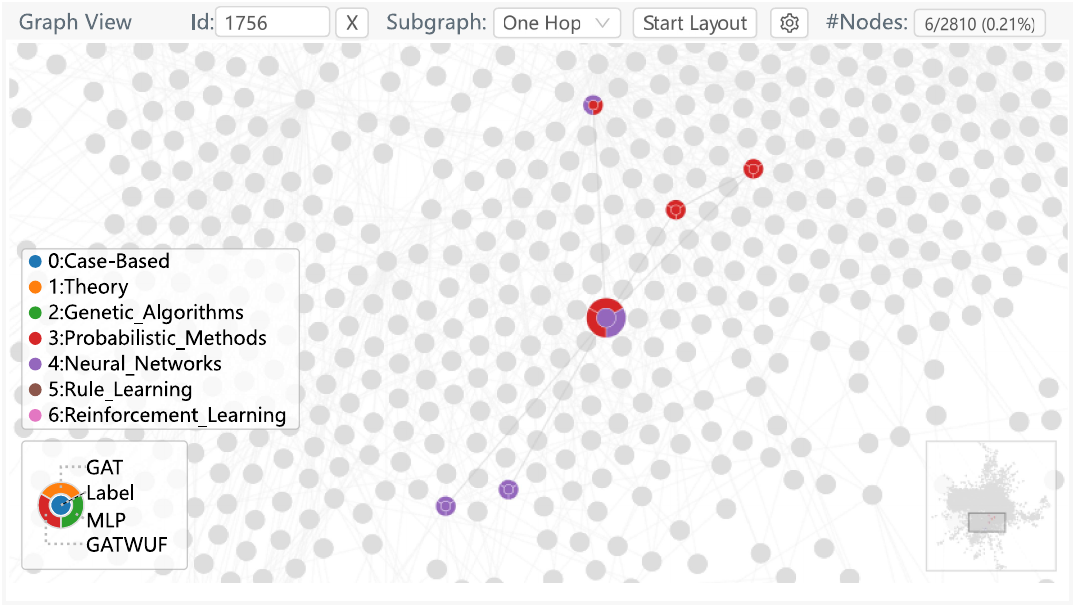}
\caption{Graph View enables users to inspect the graph structure. Node glyph in Graph View enables users to compare \textit{three model prediction results} and \textit{ground truth label} simultaneously. The color legend indicates which class the color represents.
The legend for the node glyphs shows the position at which each metric is encoded.
The color in the legend for node glyphs is only intended to show an example of node glyphs.  } 
\label{Fig.alternative_design_graph_view}
\end{figure}

\subsubsection{Graph View}

We use the classic node-link diagram with the force-directed collision-avoidance layout to visualize the graph dataset. Users can get a sense of the distribution of the selected nodes in the graph, and inspect the neighborhood of the nodes (\textbf{R2, R3}).

To further facilitate the convenient exploration of the reasons for errors, we design a node glyph to encode a group of node-level metrics.
The experts commented that they are interested in the \textit{ground truth label}, and the \textit{predictions of the GNN, GNNWUF, and MLP models}. Combining the four metrics, they are able to investigate the potential error types of the nodes. As shown in Fig.~\ref{Fig.alternative_design_graph_view}, the glyph designed to present the node-level metrics is similar to the design used in the Projection View. A legend for the glyph is also displayed at the corner of the Graph View as an easy reference for users.


The set of nodes selected in the Parallel Sets View or the Projection View is highlighted in the Graph View. Users can hover a node in the Graph View, which will be further highlighted with the radius doubled. The Graph View allows users to quickly check any interesting neighboring nodes. \mzhihua{}{Users can also switch to the ``Subgraph'' mode.
It will display the one-hop and two-hop neighbors of selected nodes,
enabling users to explore different hops of neighborhood nodes.
Users can visualize the specific subgraphs on their own and can explore them by changing the ``Subgraph'' options. For example, generated subgraphs from GNN explainability methods such as GNNExplainer~\cite{ying2019gnnexplainer} can be loaded into the system for exploration. Please check Appendix B for more details.}
An overview of the graph is displayed in the bottom right-hand corner to support users navigating the graph. Users can click the specific position in the overview to navigate the displayed area of the graph. Users can choose to filter out the unfocused nodes
to accelerate the rendering and reduce the visual clutter in the graph.
To investigate the node features and \textit{most similar features of training nodes}, users can click on the nodes of interest in the Graph View and further explore the node-level features in the Feature Matrix View. 

\begin{figure}[htb]
\centering 
\includegraphics[width=0.98\linewidth]{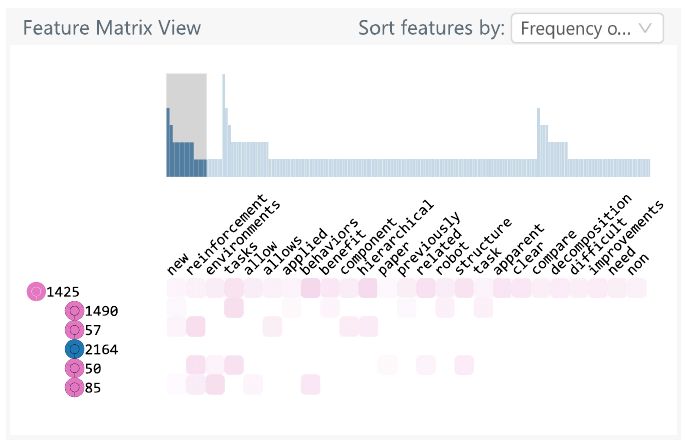}
\caption{Feature Matrix View includes brushable bar chart (Top) and feature matrix (Bottom).} 
\label{Fig.feature_matrix_View_design}
\end{figure}

\subsubsection{Feature Matrix View}

We design the Feature Matrix View to help users further explore the node features (\textbf{R3}), as shown in Fig.~\ref{Fig.feature_matrix_View_design}.  
The Feature Matrix View consists of two components, \ie, a brushable bar chart and a feature matrix.

We first assume that all the features used in our dataset range from zero to one. 
\mzhihua{}{For categorical features, one-hot encoding can be used to convert them to one-hot vectors,
where the value of each element of the vector will be either 0 and 1~\cite{hancock2020survey}. Thus, it can satisfy our assumption as well.}
The feature matrix indicates all the node features.
The color encodes the prediction label of that node and the opacity encodes the specific feature value. In the brushable
bar chart, the bar height encodes the count of any features with a value larger than 0 in the feature matrix. 
Users can brush a range of bars in the brushable bar chart and thus the feature matrix will display the specific range of feature dimensions. 
This makes it really convenient for users to inspect the features of nodes with a high dimensionality and without this design, the scalability of this view is not guaranteed. Users can change the sorting methods of feature dimensions. It can be sorted based on \textit{node ordering} or \textit{frequency of features}.
When users select a subset of nodes in Parallel Sets View and Projection View, it will display the features of selected nodes. The hierarchical clustering algorithm and optimal leaf ordering~\cite{DBLP:conf/ismb/Bar-JosephGJ01} will be employed to generate the node ordering.
After sorting the nodes, the similarity will be calculated between two consecutive nodes. If they are very similar, we highlight them by adding a border in the rectangle in the rows of corresponding nodes. 
When a node is selected in the Graph View, it will display the features of that node and the \textit{top-k most similar feature training nodes}. 
The training nodes will be sorted based on feature similarity with that node. 
When sorting feature dimensions based on the \textit{frequency of features}, the following strategy is used.
For each dimension of the features, we first count the frequency $|N|$ and then calculate the frequency of support $|SUPP|$, \ie,
the number of nodes with the same features and model prediction label as the first node.
We then calculate the support rate of the features $SUPPRATE$  by using formula: $SUPPRATE=|SUPP|/|N|$. 
Therefore, when the support rate is high, it will have a higher ranking. When the support rate between two dimensions of a feature is the same, it is sorted based on the frequency of features.
Then we can figure out what features can be supportive of the predictions for the GNN model.



\section{Evaluation}
\label{sec:evaluation}
In this section, we demonstrate the effectiveness and usability of {\name} through two case studies and structured interviews with GNN experts. We conduct two case studies with two experts E1 and E2 who have been introduced in Section~\ref{subsec:design_requirements}. 

\begin{figure}[htb]
\centering 
\includegraphics[width=0.5\textwidth]{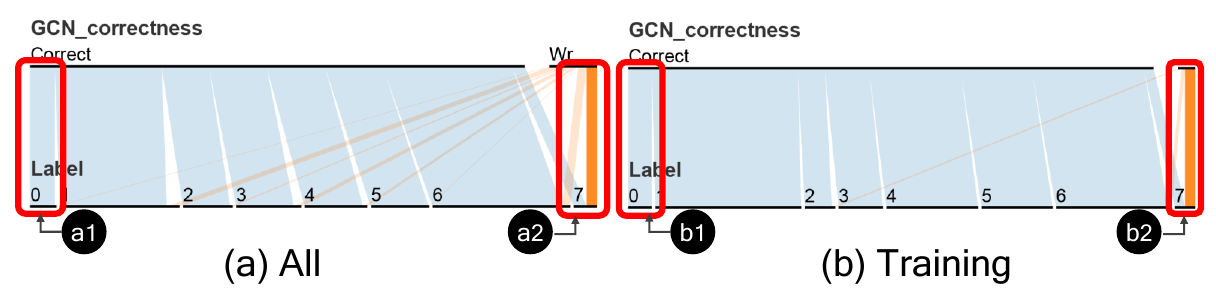}
\caption{The correlation between \textit{GCN correctness} and \textit{Label}: (a) All nodes in Amazon Photo dataset; (b) Training nodes in Amazon Photo dataset.} 
\label{Fig.case_photo_1}
\end{figure}

\begin{figure}[htb]
\centering 
\includegraphics[width=0.5\textwidth]{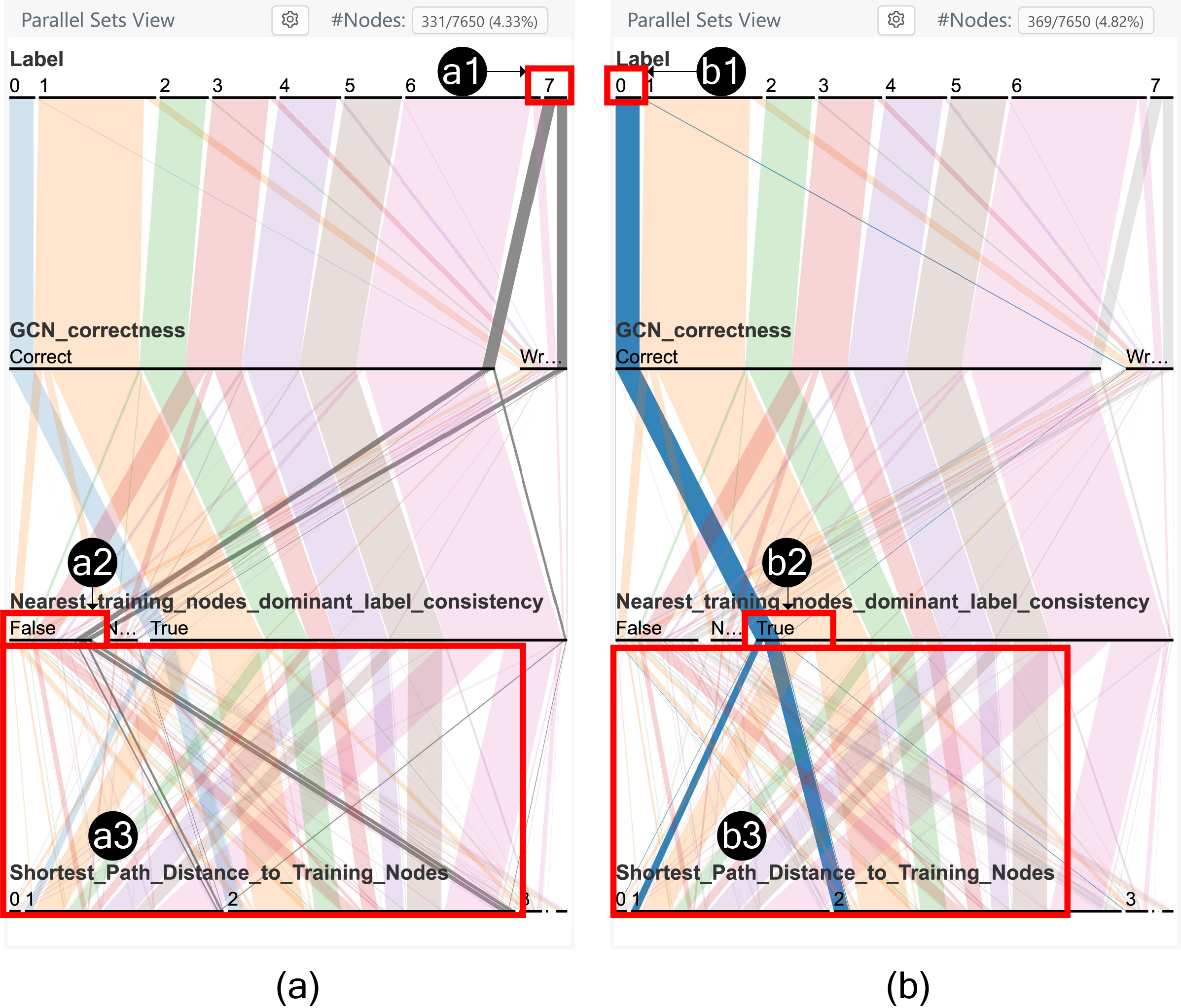}
\caption{The correlation among  \textit{Label}, \textit{GCN correctness}, \textit{nearest training nodes dominant label consistency}, and \textit{shortest path distance to training nodes}: (a) Nodes with \textit{Label} \textit{''7''}; (b) Nodes with \textit{Label} \textit{''0''}.}
\label{Fig.case_photo_2}
\end{figure}

\subsection{Case One: Error Pattern Analysis of GCN on Amazon Photo Dataset}
This case study shows how our approach helps the model researcher explore the error patterns of GCN, one of the most representative GNN models, on the Amazon Photo dataset~\cite{shchur2018pitfalls}.
The Amazon Photo dataset is a co-purchasing network of 7,650 products. In this dataset, each node represents a product and is classified into one of the eight classes, including \textit{File Photography, Digital Concepts, Binoculars \& Scopes, Lenses, Tripods \& Monopods, Video Surveillance, Lighting \& Studio,} and \textit{Flashes}. 
Each edge is a co-purchasing relationship, \ie, products are purchased by the same customer.
Each feature of the node is a vector of 0-1 value indicating whether the corresponding word appears in product reviews or not.

\subsubsection{Developing Initial Hypotheses about the Possible Error Patterns in GNN Results}
\label{sec:case_one_1}
E1 started his analysis from the Parallel Sets View. E1 found that the GCN model achieves an accuracy of 91.15\% on the whole dataset. The test accuracy is 91.80\%.
The model performance is consistent with the results reported in other papers~\cite{shchur2018pitfalls}.  
E1 changed the first axis of the Parallel Sets View to be \textit{GCN correctness} by dragging the corresponding axis to the first axis.
The total number of wrong prediction nodes is 677. 
Then, E1 explored the variables correlated with the wrong prediction.
E1 put the \textit{Label} in the second axis and found that the GCN model makes the most percentage of wrong predictions on the nodes of Class 7. 
This is indicated by the ribbon link flowing from the wrong category to the \textit{ground truth label} \textit{''7''}, which occupies the largest portion of the \textit{ground truth label} \textit{''7''} in Fig.~\ref{Fig.case_photo_1}(a2). 
E1 used the Control Panel (Fig.~\ref{Fig.system}(a)) to see the training node information by ticking \textit{``Training''}.
E1 found that the training nodes are sampled with even probability from eight classes, which is shown by a similar distribution of \textit{ground truth labels} in training nodes and all the nodes (Fig.~\ref{Fig.case_photo_1}(b)).
The number of nodes with the \textit{ground truth label} \textit{''7''} is smaller (Fig.~\ref{Fig.case_photo_1}(a2)) than the number of nodes of other labels, and the number of training nodes with the ground truth label \textit{''7''} is also small (Fig.~\ref{Fig.case_photo_1}(b2)).
Perhaps this is the reason that GCN is unable to correctly classify the nodes in Class 7.
However, E1 also found that the number of training nodes with the \textit{ground truth label} \textit{''0''} is also small (Fig.~\ref{Fig.case_photo_1}(b1)), but the GCN model correctly classifies most of the nodes in Class 0 (Fig.~\ref{Fig.case_photo_1}(a1)). E1 doubted his hypothesis and decided to further investigate the cause of wrong predictions (\textbf{R1}). 

\subsubsection{Forming the Hypothesis about Possible Error Patterns}
To analyze the error patterns found in Section~\ref{sec:case_one_1} from graph structure perspective, E1 selected four axes, including \textit{Label}, \textit{GCN correctness}, \textit{nearest training nodes dominant label consistency}, and \textit{shortest path distance to training nodes},
in the Parallel Sets View, as E1 believed that the axis \textit{Label} can help him select the specific label to check. The axis \textit{GCN correctness} can help him to filter the nodes with the correct prediction or wrong prediction. The axes \textit{nearest training nodes dominant label consistency} and \textit{shortest path distance to training nodes} help him analyze the nodes from graph structure perspective. 
After hovering over \textit{Label} \textit{''0''} (Fig.~\ref{Fig.case_photo_2}(a1)) and \textit{Label} \textit{''7''} (Fig.~\ref{Fig.case_photo_2}(b1)), E1 found that for \textit{nearest training nodes dominant label consistency}, most nodes (355/369) with \textit{Label} \textit{''0''} have a true value (Fig.~\ref{Fig.case_photo_2}(a2)), while most nodes (264/331) with \textit{Label} \textit{''7''} have a false value (Fig.~\ref{Fig.case_photo_2}(b2)).
This shows that from graph structure perspective, it is easy for nodes with \textit{Label} \textit{''0''} to find training nodes with the same \textit{ground truth label} through searching the training nodes with shortest path distance to \textit{current nodes}, while nodes with \textit{Label} \textit{''7''} cannot satisfy these conditions. Moreover, E1 can find that \textit{the shortest path distances to training nodes} for most nodes with \textit{Label} \textit{''0''} (Fig.~\ref{Fig.case_photo_2}(a3)) and \textit{Label} \textit{''7''} (Fig.~\ref{Fig.case_photo_2}(b3)) are less than or equal to 2. It means that  nearest training nodes for most nodes with \textit{Label} \textit{''0''} and \textit{Label} \textit{''7''} are located in the input scope of GCN with two layers and will have an impact on the \textit{prediction results of GCN}.
Therefore, E1 speculated that it can be the possible reason for more classification errors when GCN is applied to the nodes of \textit{Label} \textit{''7''} (\textbf{R2}).

\subsubsection{Analyzing the Cause of Error Patterns}

To further verify the cause of the error patterns identified above, E1 selected 150 wrongly-classified nodes of \textit{Label} \textit{''7''} by GCN in Parallel Sets View (Fig.~\ref{Fig.system}(b1)) and further explored other views. 
From the planes of \textit{training nodes structure influence and feature influence} in the Projection View (Fig.~\ref{Fig.system}(c)), few nodes of \textit{Label} \textit{''7''} appear in the label distribution and many prediction labels are consistent with the largest component on the right-hand side of the glyph.
\zhihua{}{It indicates that the structure and the features of training nodes of \textit{Label} \textit{‘’7’’} do not have a significant influence on the prediction results of testing nodes of \textit{Label} \textit{‘’7’’} when compared with the influence of other training nodes.
}
From the plane of \textit{surrounding node label consistency}, the \textit{label consistency} of most nodes is relatively small. 
It can be explained that the labels of the neighbors of these nodes are mostly inconsistent with the \textit{label} of the \textit{current node}.
\zhihua{}{It means that the information of neighboring nodes can confuse the GCN prediction results of the \textit{current node}.}

E1 further explored three training nodes that are also misclassified.
E1 lasso-selected them (Fig.~\ref{Fig.system}(c1)), and then selected one of the nodes in the Graph View for further checking (Fig.~\ref{Fig.system}(d1)). 
\mzhihua{}{E1 found that 
the \textit{ground truth labels} of the majority of its neighbors are different from its own \textit{ground truth label},
and its \textit{GCNWUF prediction result} is consistent with the \textit{GCN prediction result} (Fig.~\ref{Fig.system}(d1)). 
In the Feature Matrix View (Fig.~\ref{Fig.system}(e)), E1 observed that the frequency of features is very high in the front of the bar chart (Fig.~\ref{Fig.system}(e1)), and the last features of the \textit{current node} also share a lot of common features with the \textit{top-k most similar training nodes} (Fig.~\ref{Fig.system}(e2)).
It indicates that the \textit{current node} shares a lot of common features with the \textit{top-k most similar training nodes}.
Moreover, the \textit{ground truth labels} of those nodes are different from the \textit{ground truth label} of the \textit{current node} (Fig.~\ref{Fig.system}(e3)). 
Therefore, from the above observations, E1 speculated that 
the classification error of the \textit{current node} is probably due to the existence of a large number of neighboring nodes whose \textit{ground truth labels} are totally different from the \textit{current node}. 
Also, the \textit{current node} lacks discriminative features.
}

\mzhihua{}{However, E1 also spotted that the \textit{MLP prediction result} is consistent with the current node's \textit{ground truth label} (Fig.~\ref{Fig.system}(d1)). E1 quickly raised one question: \textit{Does the MLP model memorize the features of training nodes of \textit{Label} \textit{''7''} and fail to generalize to other nodes with the same \textit{ground truth labels}?} 
E1 checked the performance of MLP in the Parallel Sets View and found that MLP correctly classifies all the training nodes of \textit{Label} \textit{''7''}, but incorrectly classifies all the test nodes of \textit{Label} \textit{''7''}. 
E1 thought that it is very hard for GCN models to correctly classify the test nodes with a strong influence of the structures and the features of nodes of other \textit{classes}.
Thus, E1 believed that when training the GCN using cross-entropy loss, he can increase the loss
of the training nodes of \textit{Label} \textit{''7''} to improve the classification of
the nodes of \textit{Label} \textit{''7''}, decreasing the strong influence of nodes of other \textit{ground truth labels}~\cite{zhou2005training} (\textbf{R3}). }

\begin{figure*}[htb]
\centering 
\includegraphics[width=\textwidth]{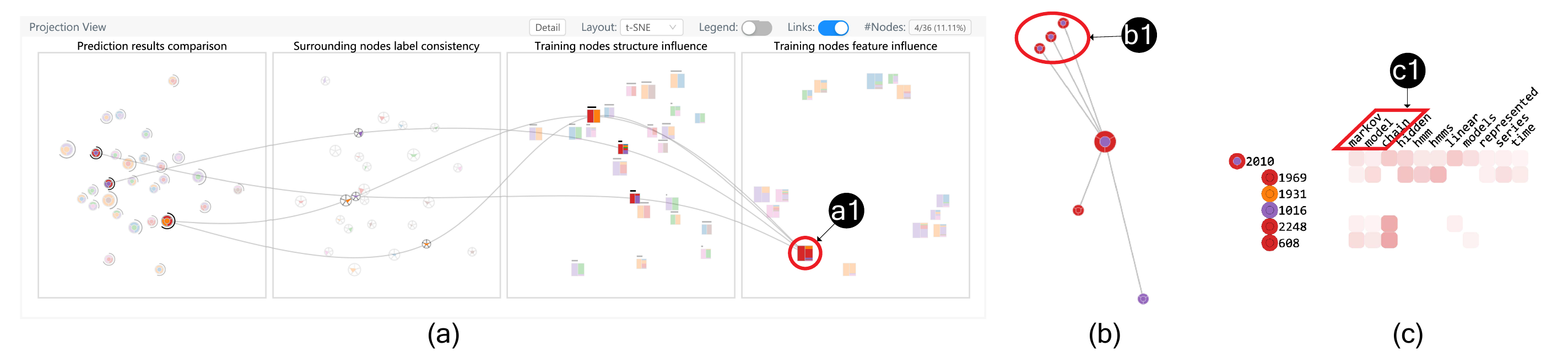}
\caption{E2 selected a cluster (a1) to inspect in Projection View (a). Then E2 selected a node in Graph View (b) to further inspect its neighborhood. E2 found that in Feature Matrix View (c), the first few words are ``markov'', ``model'', ``chain'' (c1), which are common words in the paper belong to \textit{Probabilistic Methods}. \zhihua{}{E2 verified that this node is mislabeled and should be labeled as \textit{Probabilistic Methods}.}} 
\label{Fig.case_cora_ml_1}
\end{figure*}

\subsection{Case Two: Error Pattern Analysis of GAT on Cora-ML Dataset}
The model developer, E2, often needs to use GNNs to model network data in real applications. This case study shows how {\name} assists him in analyzing another representative GNN model (\ie, the GAT model) on the Cora-ML dataset~\cite{bojchevski2017deep}.
Specifically, the Cora-ML dataset is a citation network of 2,810 scientific publications. Each node in the Cora-ML dataset represents a paper and is classified into one of seven classes, including \textit{Case-Based, Theory, Genetic Algorithms, Probabilistic Methods, Neural Networks, Rule Learning,} and \textit{Reinforcement Learning}. 
Each edge is a citation relationship. Each feature of the node is a vector, with each feature element ranging from 0 to 1. A feature element larger than 0 represents that the paper abstract contains the corresponding word.

\subsubsection{Forming the Hypothesis about Possible Error Patterns}
In the Parallel Sets View, E2 found that the GAT achieves an accuracy of 86.16\% on the whole dataset and 84.70\% on the testing set. 
To check whether there are error patterns resulting from node features rather than graph structure,
E2 selected three axes, including \textit{nearest training nodes dominant label consistency}, \textit{top-k most similar training nodes dominant label consistency}, and \textit{GAT correctness}, as E2 believed that the axis \textit{GAT correctness} can help him to filter the nodes with the correct prediction or wrong prediction. The axes \textit{nearest training nodes dominant label consistency} and \textit{top-k most similar training nodes dominant label consistency} can help him determine whether the source of influence on prediction results is the graph structure or the node features. 
E2 found an interesting set of nodes with the \textit{nearest training nodes dominant label consistency} as true, the \textit{top-k most similar training nodes dominant label consistency} as false, and the \textit{GAT correctness} as wrong. 
E2 decided to further explore them and select those nodes in the Parallel Sets View by clicking the ribbon satisfying the above conditions.

In the Projection View (Fig.~\ref{Fig.case_cora_ml_1}(a)), E2 found that in the \textit{training nodes feature influence} plane, the left side and the right side of the glyphs have mostly the same color (Fig.~\ref{Fig.case_cora_ml_1}(a1)).
This consistency means that the \textit{GAT prediction labels} are consistent with the \textit{top-k similar features training nodes dominant labels}. \zhihua{}{E2 speculated that the node features have a great impact on \textit{GAT prediction}. }
Then, E2 selected one of the clusters, and further checked the other planes of the Projection View.
In the \textit{training nodes structure influence} plane, E2 saw that the left-hand side and the right-hand side of the highlighted node glyphs are in different colors.
In the \textit{surrounding nodes label consistency} plane, it showed that the \textit{label consistency} is generally large, indicating that the surrounding \textit{ground truth label} is consistent with the current node's \textit{ground truth label}. In the \textit{prediction results comparison} plane, it was found that the \textit{prediction results of MLP} are consistent with the \textit{prediction results of GAT}. \zhihua{}{It further confirms that the node features of this cluster of nodes have a great impact on the GAT predictions on them (\textbf{R1, R2}).}

\subsubsection{Analyzing the Cause of Error Patterns}
E2 also explored the Graph View and selected a node to verify his observation.
E2 found that the node has some neighbor nodes with a different \textit{ground truth label}, as shown in Fig.~\ref{Fig.case_cora_ml_1}(b). In the Feature Matrix View (Fig.~\ref{Fig.case_cora_ml_1}(c)), E2 can see that the \textit{ground truth labels} of most training nodes are the same as its \textit{GAT prediction label}. 
The first few words are ``markov'', ``model'', ``chain'' (Fig.~\ref{Fig.case_cora_ml_1}(c1)), which are common words in papers about \textit{Probabilistic Methods}. This article has these words, but the ground truth class of this article is \textit{Neural Network}. 
\zhihua{}{E2 thought this node is mislabeled rather than misclassified. To verify whether it is labeled correctly, E2 further checked its title by hovering over the glyph in the Feature Matrix View and found that the title is \textit{“Equivalence of Linear Boltzmann Chains and Hidden Markov Models”}~\cite{mackay1996equivalence}.
After further checking the content of this article, E2 concluded that the ground truth class of this article should be \textit{Probabilistic Methods} rather than \textit{Neural Networks}. 
E2 further checked the three neighboring nodes of this node, which shares the same properties that the ground truth class is \textit{Neural Network} but all three model prediction results are \textit{Probabilistic Methods} (Fig.~\ref{Fig.case_cora_ml_1}(b1)). 
E2 found that they share the same title \textit{``Gibbs-Markov Models''}~\cite{lafferty1996gibbs} and should belong to the class \textit{``Probabilistic Methods"}. E2 suggested that the wrong labels should be identified and corrected, especially when the articles have words ``markov'', ``model'', ``chain'', and its ground truth class is \textit{Neural Network} (\textbf{R3}).}

\subsection{Expert Interviews}
\subsubsection{Participants}
To further evaluate the usefulness and usability of \name{}, we also conducted interviews with 12 experts (E3-E14, age 22-46, $\mu=26.33$, $\sigma=6.09$). 
All 12 experts have experience in the application or design of GNN models. 
None of the 12 experts are co-authors of this paper.

\subsubsection{Methodology}
We first introduced {\name} to experts and demonstrated the case study with  E1. 
After experts learned about how {\name} works, we asked them to explore {\name} by following the demonstrated workflow to find the cause of the prediction errors of individual nodes and extract general error patterns in GNN prediction results.
Finally, we asked them to finish a post-interview questionnaire with 5-point Likert scale questions (1-Strongly Disagree, 5-Strongly Agree) to collect their feedback on {\name}. 
As shown in Table~\ref{tab:quan_eval}, the questionnaire mainly comprises evaluations on the effectiveness, visual design, and usability of {\name}. 
Results and feedback are summarized accordingly.

\subsubsection{Results}
\begin{table}[]
\centering
\begin{tabular}{ c | p{6cm} | c }
\hline
 & \multicolumn{1}{c|}{Question} & Score \\ \hline
 \multicolumn{3}{c}{Effectiveness}  \\
 \hline
Q1 & The system can help me understand the overview of dataset and GNN prediction results. & 4.50 $\pm$ 0.65 \\
Q2 & The system can help me identify the error patterns of GNNs. & 4.25 $\pm$ 0.72 \\
Q3 & The system can help me analyze the causes of error patterns of GNNs. & 3.92 $\pm$ 0.76 \\ 
\hline
 \multicolumn{3}{c}{Visual design}  \\
\hline
Q4 & The Parallel Sets View is intuitive and easy to understand. & 4.08 $\pm$ 0.95 \\
Q5 & The Projection View is intuitive and easy to understand. & 3.83 $\pm$ 0.99 \\
Q6 & The Graph View is intuitive and easy to understand. & 4.50 $\pm$ 0.76 \\
Q7 & The Feature Matrix View is intuitive and easy to understand. & 4.00 $\pm$ 0.82 \\
Q8 & The overall system is intuitive and easy to understand. & 3.83 $\pm$ 0.90 \\ 
\hline
 \multicolumn{3}{c}{Usability}  \\
\hline
Q9 & It is easy to learn and use the system. & 3.67 $\pm$ 1.03 \\
Q10 & I would like to recommend this system to others who are working on diagnosing GNNs. & 4.42 $\pm$ 0.76 \\
Q11 & I think it is useful to use this system to diagnose GNNs. & 4.00 $\pm$ 0.82 \\
Q12 & I would use this system to diagnose errors of GNNs in the future. & 4.08 $\pm$ 0.76 \\ \hline
\end{tabular}
\caption{Evaluation on effectiveness (Q1-Q3), visual design (Q4-Q8), and usability (Q9-Q12) of {\name}. Scores (mean $\pm$ std) for each question are reported.}
\label{tab:quan_eval}
\end{table}

\noindent
\textbf{Effectiveness:} 
After exploring {\name}, all the experts appreciated our efforts in making such an effective system to help them understand and diagnose the GNN models. As shown in Table~\ref{tab:quan_eval}, experts agreed that {\name} can help users understand the overview of dataset and GNN prediction results, identify the error patterns, and analyze the causes of error patterns.
E5 commented that \textit{''the visualization is clear and insightful, leading to the rapid discovery of the underlying error patterns, and thereby enables future endeavors in improving the GNNs.''} 
It further supports that {\name} can help users diagnose the GNNs used in node classification tasks.

\noindent
\textbf{Visual design:} From the results on the evaluation of visual design, we observe that most of the experts (8/12) agreed that the overall visual designs of {\name} are intuitive and easy to understand. The experts agreed that the Graph View is the most intuitive and easy to understand, as shown in Table~\ref{tab:quan_eval}. In terms of the Projection View, the experts think that the glyphs can help users understand the metrics of each node but the learning curve is a bit steep.

\noindent
\textbf{Usability:} 
Most of the experts (7/12) agreed that {\name} is easy to use and easy to learn. As shown in Table~\ref{tab:quan_eval}, the experts agreed that they would like to recommend the system to others who are working on diagnosing GNNs. They also agreed that it is useful to diagnose GNNs and would like to use the system in the future. E7 further positively commented that \textit{''the response time of the system is quite fast and the overall design is clear.''} Such results support that {\name} will be widely used by users to understand and diagnose GNNs.

\noindent
\textbf{Suggestions:}
Experts also gave helpful suggestions for improving {\name} to further support their analysis of GNN models. For example, it would be more helpful if the system can provide recommendations on how to further improve GNNs if some error patterns of GNNs are found. Moreover, if the system can support customized datasets like heterogeneous graphs, the generalizability of the system will be further enhanced.

\section{Discussions and Future Work}
\label{sec:discussion}
\textbf{Generalizability:} {\name} can be applied to analyze various kinds of GNN models and different datasets. However, it can only be utilized to analyze the task of node classification. It does not support the analysis of link prediction and graph classification. Moreover, if the dataset has multiple relations or the edges have features, the system cannot be directly used to analyze those customized data. Also, users may want to
further customize some metrics to inspect the relationship between them and the model performance. However, the system does not support it currently.

\noindent
\textbf{Scalability:} 
One limitation of {\name} is its scalability and we have attempted to mitigate this issue by different means. For example,
the Projection View displays the individual node glyphs for each node. Due to the limited screen space, we cannot display up to 300 nodes. 
We improved the Projection View by using a hierarchical clustering algorithm to make it scale up to more nodes, as described in Section~\ref{sec:ProjectionView}.
The node glyphs are used to represent clusters and details can be checked on demand. This significantly improves the scalability of the Projection View. 
For the Graph View, in order to accelerate the rendering speed, we enable users to display only the focused nodes and their neighbors, without rendering other nodes. Some scalability issues cannot be easily resolved. For example, in the Projection View, if the number of labels is increasing, the number of different colors to encode different labels will also be increased. However, too many colors will increase the cognitive load for users. Moreover, as the number of selected data points is increasing, the number of links will also increase, which can result in visual clutters.




\noindent
\mzhihua{}{\textbf{Learning curve of the Projection View:} Compared with other views of {\name}, it took more time for the domain experts to understand the glyphs and metrics used in the projection view.
To further help experts understand the meaning of glyphs, glyph legends have been added to the Projection View for an easy reference. 
The Projection View plays an important role in helping users understand the properties of the selected nodes and narrow down to one or several nodes for further analysis in the Graph View and Feature Matrix View. Without this design, it is hard to explore the properties among those selected nodes.}

\noindent
\textbf{Future work:} 
In the future, we plan to generalize {\name} to other graph-related tasks, like link prediction and graph classification. 
We plan to make Parallel Sets View and Projection View more configurable, such as enabling users to define their own metrics, such as clustering coefficients to show.  We also want to further improve the running performance of {\name} and support the graph dataset with more nodes and higher dimensional features. 
Moreover, We want to generalize {\name} to support the dynamic insertion of nodes and edges to see the corresponding change in GNN prediction results.  It will help users understand how the graph structure benefits the model to improve performance.


\section{Conclusion}
\label{sec:conclusion}
In this paper, we present {\name}, a visual analytics system to help model developers and users understand and diagnose GNN model prediction results. {\name} comprises four visualization components: the Parallel Sets View enables users to see the distribution of metrics; the Projection View presents a set of 2D projections of the selected nodes according to metrics summarized from different perspectives enabling users to extract potential clusters of nodes; the Graph View shows the whole graph; and the Feature Matrix View shows the selected node feature information. It further enables users to check the detailed information of individual nodes. All four visualization components are linked together to support users to analyze GNN models simultaneously from multiple angles and extract general error patterns in GNN prediction results.
 Two case studies and expert interviews demonstrate the effectiveness and usability of our system {\name}. 


%



\ifCLASSOPTIONcompsoc
  \section*{Acknowledgments}
\else
  \section*{Acknowledgment}
\fi

We would like to thank external experts for participating in our interviews and giving us invaluable feedback. We also thank the anonymous reviewers for their detailed reviews and the DGL development team for their constructive suggestions. This research was supported in part by Hong Kong Theme-based Research Scheme grant T41-709/17N.

\ifCLASSOPTIONcaptionsoff
  \newpage
\fi



\bibliographystyle{IEEEtran}
\bibliography{main}
%


%

\begin{IEEEbiography}[{\includegraphics[width=1in,height=1.25in,clip,keepaspectratio]{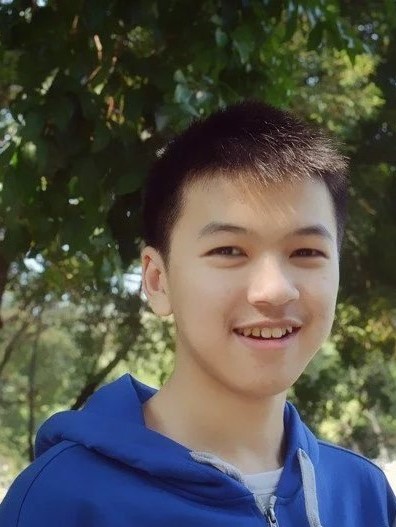}}]{Zhihua Jin} is currently a Ph.D. student at the Hong Kong University of Science and Technology (HKUST). He received his BEng degree in Computer Science and Technology from Zhejiang University in 2019. His research interests lie in the intersection of visualization and machine learning, especially explainable artificial intelligence (XAI). 
\end{IEEEbiography}

\begin{IEEEbiography}[{\includegraphics[width=1.1in,height=1.4in,clip,keepaspectratio]{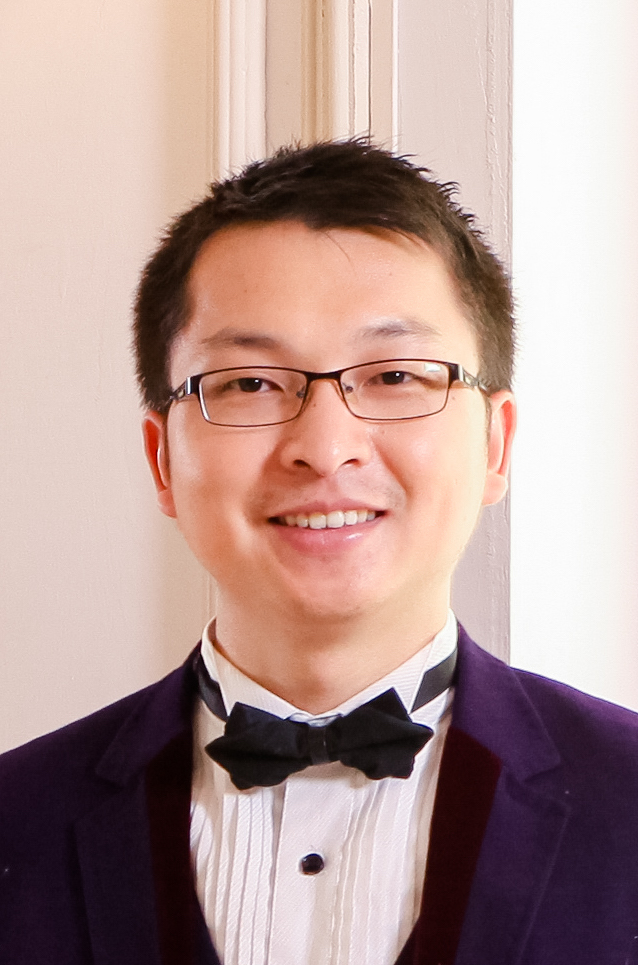}}]{Yong Wang}
is currently an assistant professor in School of Computing and Information Systems at Singapore Management University. His research interests include data visualization, visual analytics and explainable machine learning.
He obtained his Ph.D. in Computer Science
from Hong Kong University of Science and Technology in 2018. He received his B.E. and M.E. from Harbin Institute of Technology and Huazhong University of Science and Technology, respectively.
For more details, please refer to \url{http://yong-wang.org}.
\end{IEEEbiography}

\begin{IEEEbiography}[{\includegraphics[width=1in,height=1.25in,clip,keepaspectratio]{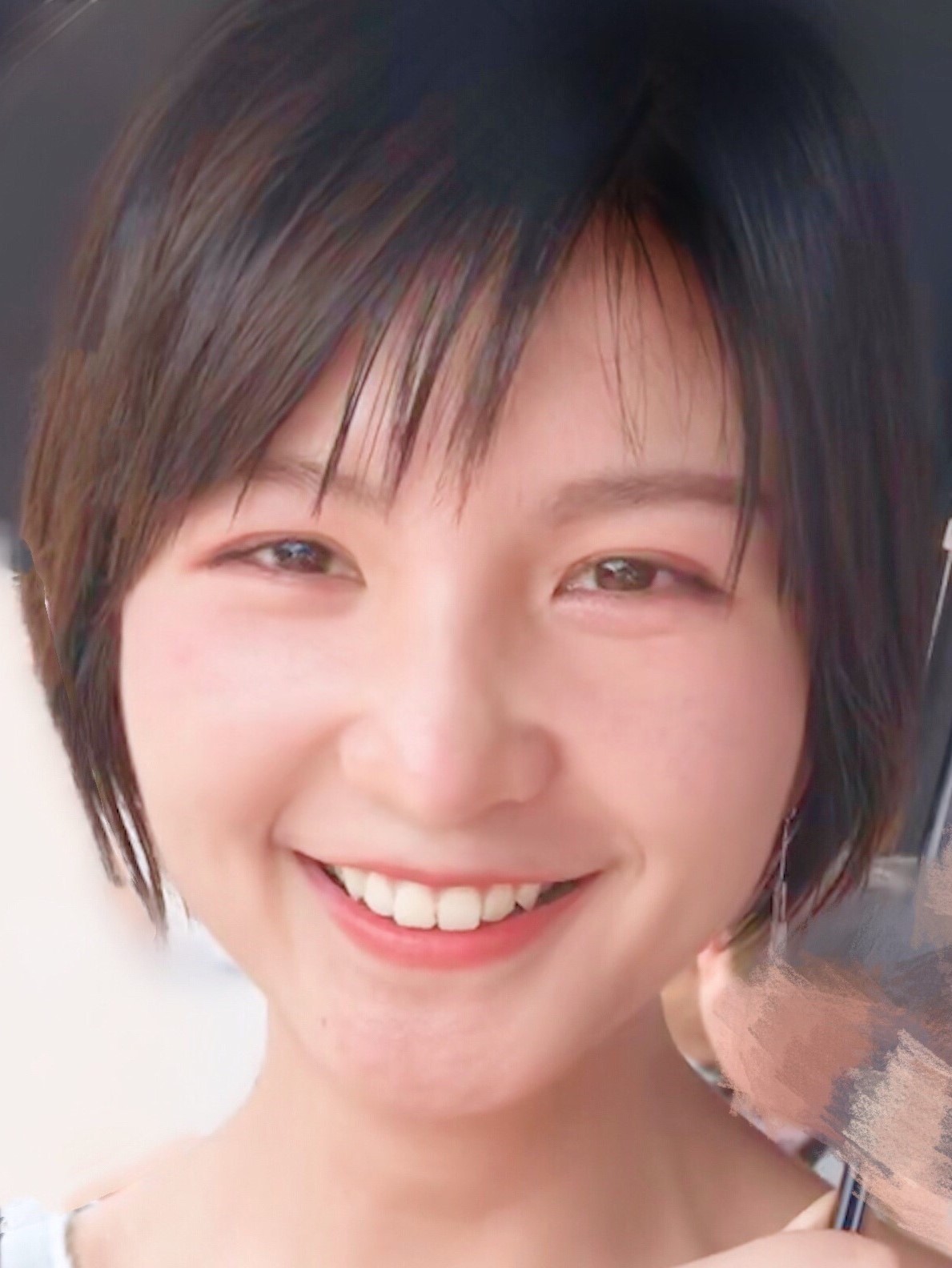}}]{Qianwen Wang}
is a PostDoc researcher at Harvard University. Her work strives to facilitate the communication between humans and machine learning models through interactive visual analytics. Her research interests include visual analytics, explainable machine learning, and narrative visualization.
She obtained her Ph.D in Hong Kong University of Science and Technology and her BS degree from Xi’an Jiaotong University. Please refer to \url{http://wangqianwen0418.github.io} for more details.
\end{IEEEbiography}

\begin{IEEEbiography}[{\includegraphics[width=1in,height=1.25in,clip,keepaspectratio]{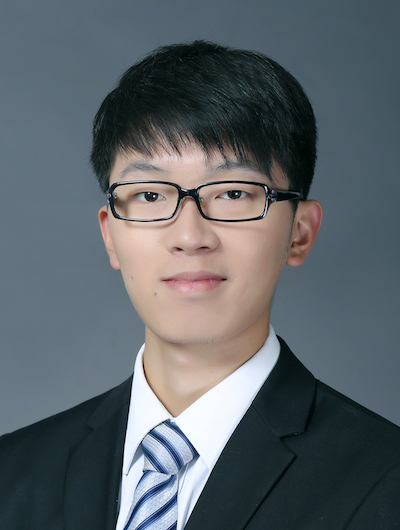}}]{Yao Ming}
is a research scientist at Bloomberg LP. His research focus on visual analytics, explainable machine learning, and natural language processing. He received a Ph.D. in Computer Science from the Hong Kong University of Science and Technology and a B.S. from Tsinghua University. For more details please refer to \url{https://www.myaooo.com}.
\end{IEEEbiography}

\begin{IEEEbiography}[{\includegraphics[width=1in,height=1.25in,clip,keepaspectratio]{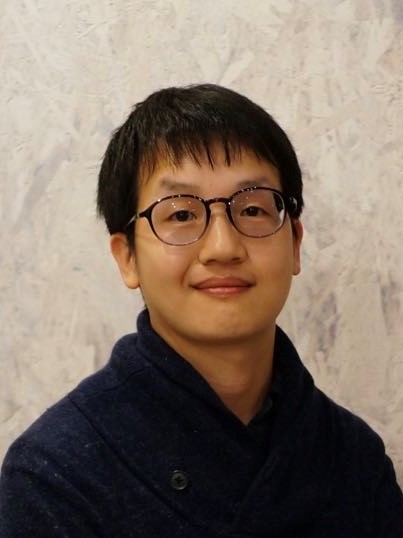}}]{Tengfei Ma} is a research staff member of IBM Research AI. Prior to joining IBM T. J. Watson Research Center, he obtained his Ph.D. from the University of Tokyo and worked as a researcher in IBM Research Tokyo for one year. Before that he got his master’s degree from Peking University and his bachelor degree from Tsinghua University. His research interests have spanned a number of different topics in machine learning and natural language processing. Particularly, his recent research is focused on graph neural networks and their applications.
\end{IEEEbiography}

\begin{IEEEbiography}[{\includegraphics[width=1in,height=1.25in,clip,keepaspectratio]{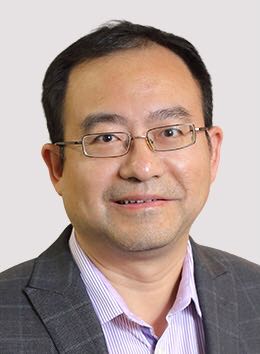}}]{Huamin Qu} is a professor in the Department of Computer Science and Engineering (CSE) at the Hong Kong University of Science and Technology (HKUST) and also the director of the interdisciplinary program office (IPO) of HKUST. He obtained a BS in Mathematics from Xi'an Jiaotong University, China, an MS and a PhD in Computer Science from the Stony Brook University. His main research interests are in visualization and human-computer interaction, with focuses on urban informatics, social network analysis, E-learning, text visualization, and explainable artificial intelligence (XAI).
\end{IEEEbiography}

\clearpage
\appendices
\section{Model Selection}
\begin{figure}[htb]
\centering 
\includegraphics[width=\linewidth]{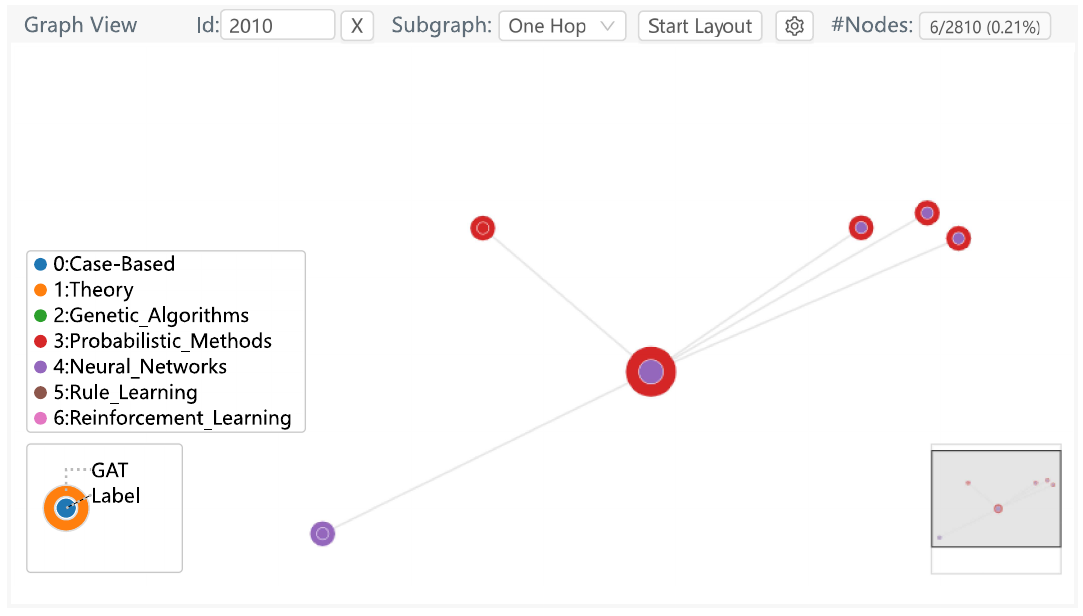}
\caption{When only selecting one model, the outer ring of the glyph encodes the prediction result of GAT. } 
\label{Fig.model_select_1}
\end{figure}

\begin{figure}[htb]
\centering 
\includegraphics[width=\linewidth]{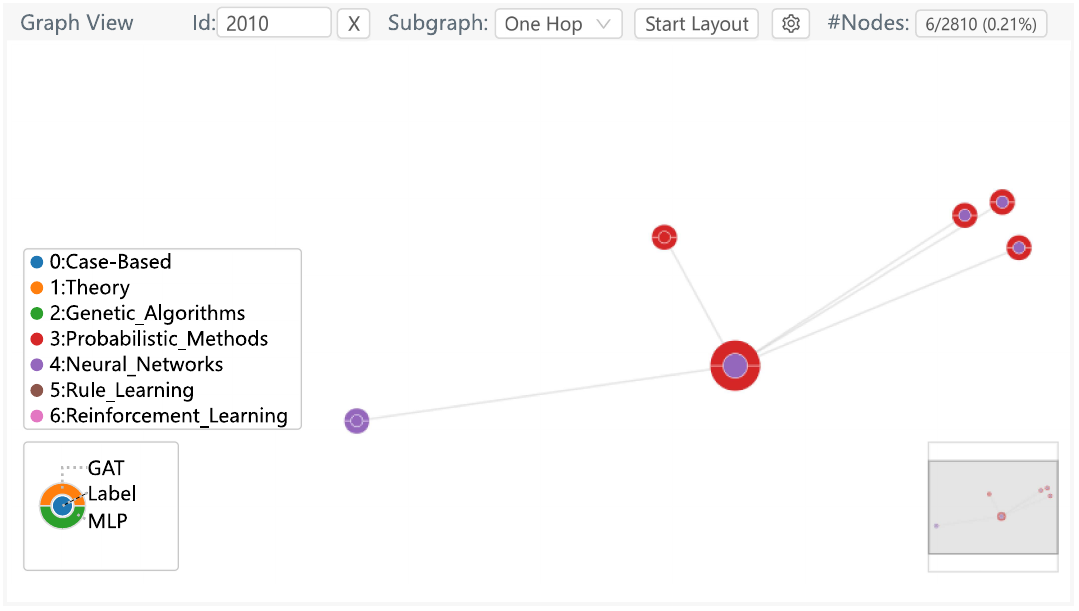}
\caption{In the Graph View, when users select two models, the outer ring of the glyph will be split into two parts. The top part encodes the prediction result of GAT and the bottom part encodes the prediction result of MLP. } 
\label{Fig.model_select_2}
\end{figure}

Users can select the inspected models in the Control Panel. It supports the simultaneous selection of one to three models.
The first model is viewed as the inspected GNN. Other models are considered proxy models. If users only select one or two models to inspect, the node glyph design and distance function will change accordingly. 
The outer ring of the glyph will be evenly split to visualize the prediction results of the respective models.
It will first display the prediction result of the first model at the top, and then clockwise display the prediction result of the other models, as shown in Fig.~\ref{Fig.model_select_1} and Fig.~\ref{Fig.model_select_2}.
The distance function in the Prediction Results Comparison plane of the Projection View will only take into account the prediction results of the currently selected models.

\section{Visualizing the Subgraphs from GNN Explainability Methods}
\begin{figure}[htb]
\centering 
\includegraphics[width=\linewidth]{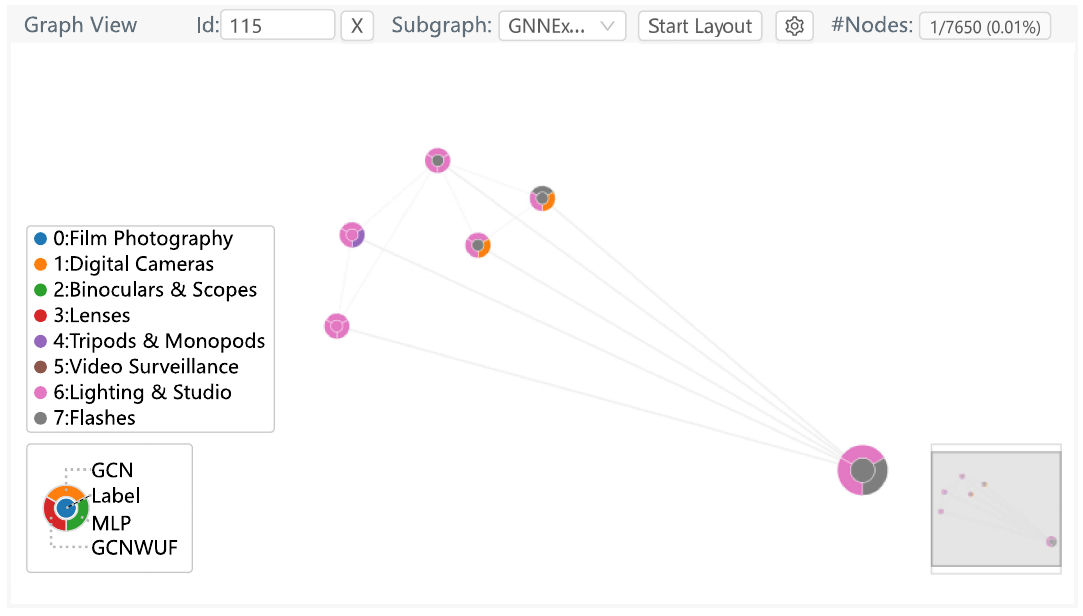}
\caption{The Graph View shows the subgraph generated from GNNExplainer. It is applied to explain the GCN prediction result of the Node 115 in the Amazon Photo dataset. } 
\label{Fig.gnnexplainer}
\end{figure}

\begin{figure}[htb]
\centering 
\includegraphics[width=\linewidth]{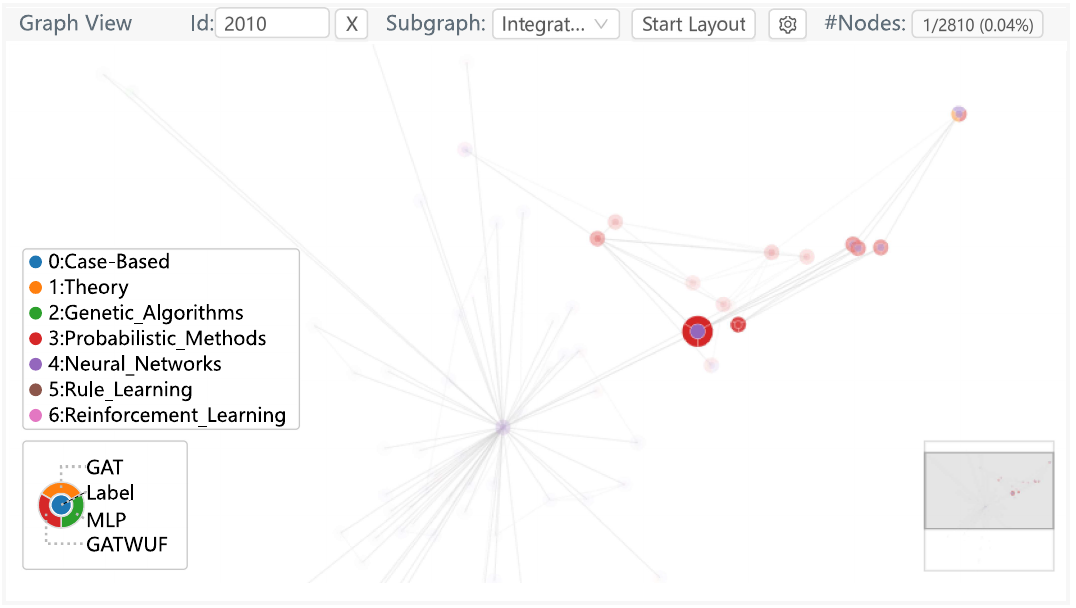}
\caption{The attribution values from Integrated Gradients method are encoded by the opacity of the node glyph. The Integrated Gradients method is applied to explain the GAT prediction result of Node 2010 in the Cora-ML dataset. } 
\label{Fig.integratedgradients}
\end{figure}

GNN explainability techniques typically generate a subgraph to explain model predictions, such as GNNExplainer (Fig.~\ref{Fig.gnnexplainer}), or calculate attribution values for each node to explain model predictions, such as Integrated Gradients (Fig.~\ref{Fig.integratedgradients}).
Users can upload the results of these methods to the backend and visualize them in the Graph View.
More precisely, users need to collect information about nodes, edges, and node weights that can be visualized in the Graph View.
In the Graph View, users can change the Subgraph options to select the methods. 
We use the opacity of the node's glyph to encode the node's weight. If the node weight is 1, then the opacity is 1. If the node weight is 0, then the opacity is 0. The results from these methods also support conclusions in the case study. For example, from Fig.~\ref{Fig.gnnexplainer}, it can be observed that the subgraph contains some nodes with different ground truth labels, which supports the conclusion that the neighboring nodes with different ground truth labels have a strong effect on the prediction of GCN in Case One. From the Integrated Gradients results (Fig.~\ref{Fig.integratedgradients}), we find that the prediction of the Node 2010 attributed to many nodes with class \textit{“Probabilistic Methods”}. It is also in line with observation in Case Two, where the neighbors of Node 2010 have some nodes with class \textit{“Probabilistic Methods”} and have an influence on the prediction result of GAT.

\end{document}